\documentclass[sigconf]{acmart}

\usepackage{tabularx}
\usepackage{enumitem}
\usepackage{subcaption}
\usepackage{algorithm}
\usepackage{algorithmicx}
\usepackage{algpseudocode}
\usepackage{mathtools}
\usepackage{diagbox}
\usepackage{svg}
\algtext*{EndIf} 
\algtext*{EndFor} 
\algtext*{EndWhile}

\AtBeginDocument{%
  \providecommand\BibTeX{{%
    \normalfont B\kern-0.5em{\scshape i\kern-0.25em b}\kern-0.8em\TeX}}}

\setcopyright{acmcopyright}
\copyrightyear{2018}
\acmYear{2018}
\acmDOI{XXXXXXX.XXXXXXX}

\acmConference[SIGMOD '25]{In Proceedings of the 2025 International Conference on Management of Data}{June 22--27, 2025}{Berlin, Germany}
\acmPrice{15.00}
\acmISBN{978-1-4503-XXXX-X/18/06}

\newcommand*\justify{%
  \fontdimen2\font=0.4em
  \fontdimen3\font=0.2em
  \fontdimen4\font=0.1em
  \fontdimen7\font=0.1em
  \hyphenchar\font=`\-
}

\renewcommand{\texttt}[1]{%
  \begingroup
  \ttfamily
  \begingroup\lccode`~=`/\lowercase{\endgroup\def~}{/\discretionary{}{}{}}%
  \begingroup\lccode`~=`[\lowercase{\endgroup\def~}{[\discretionary{}{}{}}%
  \begingroup\lccode`~=`.\lowercase{\endgroup\def~}{.\discretionary{}{}{}}%
  \catcode`/=\active\catcode`[=\active\catcode`.=\active
  \justify\scantokens{#1\noexpand}%
  \endgroup
}

\begin{document}


\title[SymphonyQG: Towards Symphonious Integration of Quantization and Graph for Appr. Nearest Neighbor Search]{SymphonyQG: Towards Symphonious Integration of Quantization and Graph for Approximate Nearest Neighbor Search}


\author[1]{Yutong Gou\textsuperscript{\dag}}
\affiliation{%
  \institution{Nanyang Technological University}
  \country{Singapore}}
\email{yutong003@e.ntu.edu.sg}

\author[2]{Jianyang Gao\textsuperscript{\dag}}
\affiliation{%
  \institution{Nanyang Technological University}
  \country{Singapore}}
\email{jianyang.gao@ntu.edu.sg}

\author[3]{Yuexuan Xu}
\affiliation{%
  \institution{Nanyang Technological University}
  \country{Singapore}}
\email{yuexuan001@e.ntu.edu.sg}

\author[4]{Cheng Long\textsuperscript{*}}
\affiliation{%
  \institution{Nanyang Technological University}
  \country{Singapore}}
\email{c.long@ntu.edu.sg}

\date{\today}
\thanks{\textsuperscript{\dag}Equally contributed to this work.}
\thanks{\textsuperscript{*}Corresponding author.}

\renewcommand{\shortauthors}{Gou and Gao, et al.}

\begin{abstract}
Approximate nearest neighbor (ANN) search in high-dimensional Euclidean space has a broad range of applications.
Among existing ANN algorithms, graph-based methods have shown superior performance 
in terms of the time-accuracy trade-off.
However, they face performance bottlenecks due to the
random memory accesses caused by the searching process on the graph indices and 
the costs of computing exact distances to guide the searching process. 
To relieve the bottlenecks, a recent method named NGT-QG makes an attempt by integrating quantization and graph.
It (1) replicates and stores the quantization codes of a vertex's neighbors compactly so that they can be accessed sequentially,
and (2) uses a SIMD-based implementation named FastScan to efficiently estimate distances
based on the quantization codes in batch for guiding the searching process. 
While NGT-QG achieves promising improvements over the vanilla graph-based methods, it 
has not fully unleashed the potential of integrating quantization and graph.
For instance, it entails a re-ranking step to compute exact distances at the end, which introduces extra random memory accesses;
its graph structure is not jointly designed considering the in-batch nature of FastScan, which causes wastes of computation in searching.
In this work, following NGT-QG, we present a new method named SymphonyQG, 
which achieves more symphonious integration of quantization and graph (e.g., it avoids the explicit re-ranking step and refines the graph structure to be more aligned with FastScan).
Based on extensive experiments on real-world datasets, SymphonyQG establishes the new state-of-the-art in terms of the time-accuracy trade-off:
at 95\% recall, SymphonyQG achieves 1.5x-4.5x QPS compared with the most competitive baselines 
and achieves 3.5x-17x QPS compared with the classical library HNSWlib across all tested datasets. 
At the same time, its indexing is at least 8x faster than NGT-QG.
\end{abstract}


\maketitle
\sloppy

\section{Introduction}
\label{sec: intro}

Nearest neighbor search on high-dimensional vector data is a fundamental problem and has
many real-world applications, including
databases~\cite{wang_survey,mohoney2023high,wang2021milvus, single_store}, 
information retrieval~\cite{asai2023retrieval,colbert,colbertv2,mohoney2023high}, 
large language models~\cite{li2022survey,lewis2020retrieval,chang2020pre,kim2022applications} 
and recommendation system~\cite{schafer2007collaborative,wei2020analyticdb,wang2023integrity}.
Since the cost of searching for exact nearest neighbors is expensive
when the database of vectors is large, researchers and practitioners usually aim to develop algorithms for approximate nearest neighbors (ANN) search, so as to achieve a better accuracy-efficiency trade-off.

Many methods have been developed for ANN query~\cite{fanng, nsw, fu2016efanna, iwasaki2016pruned, spann, rng, hnsw,nsg,2019diskann, SPTAG, nssg, tau-mng, azizi2023elpis, pq, pqfs, opq, addq, guo2020accelerating, babenko2014inverted, gan2012locality, datar2004locality, huang2015query, indyk1998approximate, sun2014srs, tao2010efficient, beygelzimer2006cover, arya1993approximate, gu2022parallel, mtree,wang2024distancecomparisonoperatorsapproximate,yang2024bridgingspeedaccuracyapproximate, roar_graph, han2024efficient} (a more detailed review will be presented in Section~\ref{sec: related}).
Among them, 
graph-based methods show superior search performance, 
which has been widely observed in benchmarks and competitions~\cite{graph_benchmark, li2019approximate, dobson2023scaling_billion_benchmark, annbenchmark, bigann2021, bigann2023}.
In these methods, each vector in the database is viewed as a vertex in the graph. 
During indexing, these methods build a graph-based index by adding edges among the vertices.
When a query comes, these methods usually adopt a greedy beam search algorithm~\cite{hnsw, nsg, 2019diskann, nssg} to find NN (more details  will be provided in Section~\ref{subsec:ann and graph}).

Nevertheless, 
graph-based methods usually face performance bottlenecks in both memory and CPU during querying.
To be more specific,
in each iteration of the search process,
these methods require to access raw data vectors of the current vertex's neighbors in order to compute their
exact distances from the query vector~\cite{hnsw, nsg, 2019diskann, fu2016efanna, tau-mng}.
This process is costly because 
(1) it would incur many random memory accesses since these vectors are stored separately in memory (please refer to Figure~\ref{fig:main}(a)), which is unfriendly to cache,
and (2) computing exact distances between high-dimensional vectors frequently is expensive.

\begin{figure}[htbp]
    \centering
    \includegraphics[width=\linewidth]{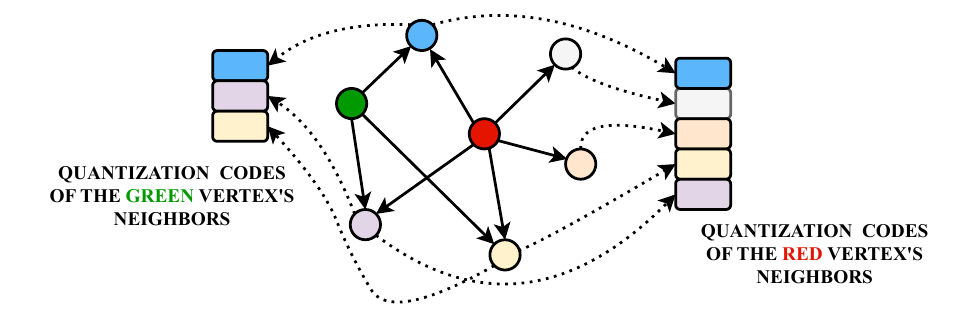}
    \vspace{-4mm}
    \caption{
    An illustration of data layout.
    As illustrated, quantization codes of the red vertex's neighbors are packed and stored together at the 
    red vertex's side. 
    The same applies to the green vertex.
    }
    \Description[<short description>]{}
    \label{fig:batch_graph}
    \vspace{-2mm}
\end{figure}

To boost the search performance of vanilla graph-based indices, a method called NGT-QG is proposed in the open-source library NGT by Yahoo Japan~\cite{ngtlib}.
NGT-QG proposes a novel framework, namely \textit{QG}, by integrating \emph{quantization}~\cite{pq} and a graph-based index for reducing the aforementioned two types of cost as follows.
(1) To mitigate random memory accesses, NGT-QG compresses raw vectors into short codes (namely quantization codes)
using product quantization (PQ)~\cite{pq}.
And for each vertex, it duplicates and stores the quantization codes of its neighbors compactly on its side in main memory (see Figure~\ref{fig:batch_graph}).
Recall that during the search process of vanilla graph-based methods,
the algorithm uses raw data vectors to compute 
exact distances between the current vertex's neighbors and the query. 
In NGT-QG, this process  is replaced by using the quantization codes of the vertex's neighbors for estimating their distances.
Since these quantization codes are stored compactly,
the memory is accessed in a sequential pattern, thus avoiding random memory accesses in each search iteration.
(2) To avoid the costs of computing the exact distances, NGT-QG estimates distances using FastScan~\cite{pqfs, quickadc}. 
Specifically, it packs every 32 quantization codes in a batch 
(i.e., the compactly stored quantization codes) and re-organizes the data layout.
In this way, during searching, it can estimate distances for 32 data vectors simultaneously via a series of SIMD-based operations, which has been shown to be highly efficient~\cite{pqfs}. 
According to the well-acknowledged ANN-Benchmark~\cite{annbenchmark}, 
NGT-QG is one of the algorithms that achieve the state-of-the-art performance in terms of time-accuracy trade-off. 
For example, when reaching 95\% recall on the SIFT dataset, it has been reported to be 2.4x faster than the classical open-source library HNSWlib~\cite{hnswlib}.

Despite the promising search performance of NGT-QG, the potential of integrating quantization and graph-based indices
has not been fully unleashed,
and we observe the following limitations of NGT-QG in querying and indexing.
\underline{\textbf{Querying}}:
The limitation in querying lies in its quantization method and search algorithm.
Specifically, the quantization method used in NGT-QG (i.e., PQ) has no guarantee on the accuracy of its estimated distances and has been observed to fail disastrously on some real-world datasets~\cite{gao2024rabitq} 
(also observed in our experiments in Section~\ref{sec: experiments}).
This is highly undesirable because ANN is used as an infrastructure in support of various downstream applications. 
The disastrous failure of ANN could cause unforeseen accidents in production environments.
As for the search algorithm, recall that NGT-QG uses estimated distances when searching on a graph. 
This entails a re-ranking step to achieve high recall.
To be more specific, at the end of the search process,
it requires to compute exact distances of selected candidates (i.e., vertices with small estimated distances) with their raw data vectors.
Subsequently, candidates with the smallest exact distances are returned as the final result for the given query.
The re-ranking process introduces extra random memory accesses
since it needs to access raw data vectors that are not stored compactly.

\underline{\textbf{Indexing}}:
The limitation in indexing lies in the efficiency of indexing and the graph structure.
On the one hand, NGT-QG only uses FastScan for accelerating the search algorithm and still incurs long time for indexing.
On the other hand,
the graph structure is not fully aligned with the search algorithm.
Recall that NGT-QG adopts FastScan to estimate distances for each vertex's neighbors, 
and FastScan always estimates a batch of distances simultaneously~\cite{quickadc} by SIMD-based operations.
When the number of a vertex's neighbors
is not a multiple of the batch size, FastScan would estimate a non-full batch of distances
and waste some computation that could have been used
for estimating distances for more vectors.
For example, in the graph of NGT-QG on Deep-1M dataset~\footnote{We build the index of NGT-QG according to parameters provided in ANN-Bechmarks~\cite{annbenchmark}. 
More details can be found in Section~\ref{sec: experiments}.}, about 23\% of vertices have non-full batches.
In summary, the limitation of NGT-QG includes both the issues of its quantization technique and the issue that graph-based indices and quantization have not been fully symphoniously integrated.

In this paper, following NGT-QG, we develop a new method named SymphonyQG which 
targets more \underline{symphon}ious integration of a superior \underline{q}uantization technique and \underline{g}raph-based indices for ANN query.
Specifically, SymphonyQG is built upon four major ideas.
(1) We incorporate
the RaBitQ~\cite{gao2024rabitq} quantization method, which is recently available in the literature, to the scheme of searching graph-based indices with FastScan.
RaBitQ is superior over PQ in that it
provides an unbiased estimator with rigorous error bounds and is shown to have better practical performance~\cite{gao2024rabitq}.
Nevertheless, RaBitQ is originally used with IVF~\cite{gao2024rabitq} and has not been used with graphs. 
We solve some technical challenges such as those in normalizing the vectors and in efficiently estimating distances 
with FastScan when searching on graphs.
(2)
We jointly design a new search algorithm and a specialized data layout to further reduce random memory accesses. 
Specifically, our method stores the raw data vectors compactly with the data needed in graph-based searching (e.g., the neighbors and their quantization codes).
When the method visits a vertex during searching, it computes its exact distances based on the raw vectors and updates the NN (i.e., implicitly conducting re-ranking).
This removes the explicit re-ranking step which incurs random memory accesses. 
In addition, a new search algorithm is designed, which uses multiple estimated distances to guide searching so that the true NN is less likely to be missed in the aforementioned implicit re-ranking.
(3) We propose to leverage FastScan to accelerate the process of building a graph-based index. 
Specifically, we build a graph-based index by starting with an initialized graph and then adjusting the graph iteratively. 
In each iteration, we use our search algorithm (which is equipped with FastScan) for finding nearest neighbors of a vertex as candidates to be linked from the vertex in the graph.
According to our experimental results, our index construction method is at least  8x faster than NGT-QG on real-world datasets.
(4) 
We propose a graph refinement strategy which supplements edges for the graph
so that the graph-based index 
surely has the out-degree of every vertex to be a multiple of the batch size.
Thus, the refined graph is better aligned with FastScan 
and the in-batch computation of FastScan is fully utilized.

In summary, this paper makes the following contributions:
\begin{enumerate}
\item \textbf{Novel ANN Algorithm}. 
We propose SymphonyQG, an algorithm which achieves more symphonious integration of quantization and graph-based indices.
Compared with NGT-QG, our symphonious integration brings superior and more stable performance on ANN query and requires significantly shorter time in indexing.

\item \textbf{Open-Source Library~\footnote{\url{https://github.com/gouyt13/SymphonyQG}}}.
Based on SymphonyQG, we provide a well-optimized open-source library for ANN query.
The library provides both header-only implementations in C++ and convenient APIs in Python.
Therefore, the research community  and practitioners can easily evaluate and utilize our library with a few lines of Python codes.

\item \textbf{Extensive Experiments}.
We conduct extensive empirical studies on real-world datasets. The results verify that 
SymphonyQG establishes the new state-of-the-art in terms of the time-accuracy trade-off.
At 95\% recall ($K=10$), SymphonyQG achieves 1.5x-4.5x QPS compared with the most competitive baselines and achieves 3.5x-17x QPS compared with the classical library HNSWlib across all tested datasets. 
The experiments conducted on two datasets with 100 million vectors verify its scalability. 
The ablation studies verify the effectiveness of each individual proposed technique.

\end{enumerate}

The remainder of this paper is organized as follows.
Section~\ref{sec: prelimin} reviews graph-based ANN methods, FastScan and RaBitQ.
Section~\ref{sec: method} presents the SymphonyQG method.
Section~\ref{sec: experiments} reports extensive experimental results on real-world datasets.
Section~\ref{sec: related} discusses related work.
Section~\ref{sec: conclusion} provides conclusions and discussions.

\section{Preliminaries}
\label{sec: prelimin}
In this section, we briefly introduce the ANN query and the general workflow of graph-based methods in Section~\ref{subsec:ann and graph}. To make the paper self-contained, we present a quick revisit to RaBitQ and FastScan in Section~\ref{subsec: rabitq}.

\subsection{ANN and Graph-based Methods}
\label{subsec:ann and graph}

\smallskip\noindent\textbf{Approximate Nearest Neighbor Search.}
Given a database in which each object is a high-dimensional vector, nearest neighbor (NN) query targets to find the vector that has the smallest distance to a query vector.
In practice, the latency of finding exact NN is unacceptable due to the increasing size of the database and curse of dimensionality~\cite{indyk1998approximate}.
Thus, researchers have developed approximate nearest neighbor (ANN) search algorithms for achieving better trade-offs between accuracy and efficiency.
In this paper, we focus on in-memory ANN query, where data vectors and indices are all stored in RAM.
Besides, ANN query often targets to find $K$ approximate nearest neighbors. 
When describing algorithms, we omit $K$ for the convenience of presentation, while all methods developed in this paper are applicable to a general $K$.

\smallskip\noindent\textbf{Graph-based ANN Algorithms.}
Many graph-based indices have been proposed for ANN query~\cite{fanng, nsw, fu2016efanna, iwasaki2016pruned, spann, rng, graph_benchmark, hnsw,nsg,2019diskann,li2019approximate, SPTAG, nssg, tau-mng, azizi2023elpis}.
These methods produce the state-of-the-art search performance because they can usually find nearest neighbors by considering only a small number of vectors as candidates through graph-based searching~\cite{hcnng}.
Specifically, before querying, 
these methods build a graph on the database, where each data vector corresponds to a vertex in the graph.
When a query comes, an algorithm named greedy beam search~\cite{hnsw, nsg, 2019diskann, nssg} is usually adopted.
During searching, the algorithm maintains a set named the beam set which has the size named the beam size of up to $n_b$.
The search process starts by adding an entry point
to the beam set and proceeds iteratively.
In each iteration, the algorithm finds the vector which is nearest to the query vector in the beam set and has not been visited so far.
Then, the algorithm marks it as visited and computes the exact distances between the query vector and all its neighbors on the graph.
The neighbors are next inserted into the beam set.
If the beam set has more than $n_b$ objects after insertion, then only the nearest $n_b$ vectors would be kept.
The algorithm terminates when all $n_b$ vectors in the beam set are marked as visited. 
The nearest vector in the beam set is then returned as the answer.
It is clear that a larger beam size indicates that more vectors  would be visited during searching, and thus, corresponds to higher accuracy and larger latency.

\subsection{RaBitQ and FastScan}
\label{subsec: rabitq}
Quantization is a family of methods which compress high-dimensional vectors into short quantization codes~\cite{pq, pqfs, opq, addq, guo2020accelerating, babenko2014inverted, gao2024rabitq}.
During querying, they can utilize the short quantization code of a data vector to estimate the distance between the data vector and a query vector.
A recent study~\cite{gao2024rabitq} proposes a new quantization method called RaBitQ.
Unlike PQ and its variants~\cite{pq, pqfs, opq, addq, guo2020accelerating, babenko2014inverted}, RaBitQ provides an unbiased estimator with a theoretical error bound for the distances. In addition, it can 
use shorter quantization codes to produce consistently better empirical accuracy than PQ and its variants on real-world datasets. Thus, we consider using RaBitQ in our method.

Specifically, RaBitQ first reduces the question of estimating distances between two vectors to that of estimating inner product between their normalized vectors, which is restated in the following equations.
\begin{align}  
    &\|\mathbf{o}_r-\mathbf{q}_r \| ^2 =\| (\mathbf{o}_r-\mathbf{c}) - (\mathbf{q}_r-\mathbf{c})\| ^2 
    \\ = &\|\mathbf{o}_r-\mathbf{c}\| ^2+\|\mathbf{q}_r-\mathbf{c}\| ^ 2 - 2\|\mathbf{o}_r-\mathbf{c}\| \cdot \|\mathbf{q}_r-\mathbf{c}\|\cdot 
        \langle \mathbf{o,q} \rangle 
        \label{eq: normalization}
\end{align}
Here,
$\mathbf{o}_r$ and $\mathbf{q}_r$ denote the data vector and the query vector, respectively;
$\mathbf{c}$ denotes a vector used for normalizing the data vectors;
$\mathbf{q} := \frac{\mathbf{q}_r-\mathbf{c}}{\|\mathbf{q}_r-\mathbf{c}\|}$ denotes the normalized query vector; 
and $\mathbf{o} := \frac{\mathbf{o}_r-\mathbf{c}}{\|\mathbf{o}_r-\mathbf{c}\|}$ denotes the normalized data vector.
In \cite{gao2024rabitq}, RaBitQ is used in combination with the IVF index, and uses the centroid of each cluster in IVF as the vector $\mathbf{c}$ for normalization.
Therefore, in Equation~(\ref{eq: normalization}), $\|\mathbf{o}_r-\mathbf{c}\|$ can be pre-computed during index construction; 
and
$\|\mathbf{q}_r-\mathbf{c}\|$ can be computed once for each cluster and shared by all data vectors in the same cluster.
Thus, the question is reduced to efficiently  estimating
the inner product of two unit vectors $ \left< \mathbf{q,o} \right>$. 

During the index phase, RaBitQ considers a set of randomly rotated bi-valued unit vectors as the quantization codebook, which is restated as follows.
\begin{align}
\label{eq: codebook rabitq}
    \mathcal{C}_{rand} := \left\{ P \mathbf{x} \ \big| \  \mathbf{x}[i]  \in \left\{ + \frac{1}{\sqrt {D} }, - \frac{1}{\sqrt {D} }  \right\}, i=1,2,3,...,D  \right\} 
\end{align}
Here, $P$ is a random orthogonal matrix~\cite{johnson1984extensions} (a type of Johnson-Lindenstrauss Transformation) and $\mathbf{x}[i]$ denotes the $i$th coordinate of the vector $\mathbf{x}$. 
Let $\mathbf{\bar o}$ be the nearest vector in the codebook of the data vector $\mathbf{o}$ (i.e., $\mathbf{\bar o}$ is the quantized vector of $\mathbf{o}$).
Because the vector $\mathbf{\bar o}$ in the codebook corresponds to a bi-valued vector $\mathbf{\bar x}$, i.e., $\mathbf{\bar o}=P \mathbf{\bar x}$, 
the quantization code can be represented as a $D$-bit string, with 1 bit for each dimension.

During the query phase, 
it computes an estimator $\left< \mathbf{\bar o}, \mathbf{q}\right>/\left< \mathbf{\bar o}, \mathbf{o}\right>$ for unbiasedly estimating $\left< \mathbf{o,q}\right>$.
It is noted that $\left< \mathbf{\bar o}, \mathbf{o}\right>$ is independent of the query and can be pre-computed in the index phase. 
$\left< \mathbf{\bar o}, \mathbf{q}\right>$ can be efficiently computed for a batch of quantization codes via FastScan~\cite{pqfs, quickadc}. 
Specifically, the computation of $\left< \mathbf{\bar o}, \mathbf{q}\right>$ 
is equivalent to that of the inner product between the two vectors after they are \emph{reversely} rotated 
(i.e., via multiplying the vectors by $P^{-1}$). The following equation provides the deduction of the equivalence.
\begin{align}
\left< \mathbf{\bar o}, \mathbf{q}\right>=\left< P\mathbf{\bar x}, \mathbf{q}\right>=\left< \mathbf{\bar x}, P^{-1}\mathbf{q}\right>  
    \label{eq: rotation}
\end{align}
FastScan~\footnote{
Our implementation of FastScan is largely based on that of Faiss. 
For more details of the implementation, please refer to the Faiss wiki~\cite{faiss_fastscan}.}
can be used to compute $\left< \mathbf{\bar x}, P^{-1}\mathbf{q}\right>$ as follows.
It splits the query vector $\mathbf{q}$ into $D/4$ sub-segments where each sub-segment has $4$ dimensions. 
Since $\mathbf{\bar x}$ is a bi-valued vector,
the result of inner product $\left< \mathbf{\bar x}, P^{-1}\mathbf{q}\right>$ in each 4-dimensional sub-segment has up to $2^4$ different values. 
It pre-computes the $2^4$ values and forms a look-up-table (LUT). 
Thus, the inner product for a particular $\mathbf{\bar x}$ can be computed by looking up the LUTs for all the $D/4$ sub-segments.
FastScan~\cite{quickadc} proposes to host the LUTs in SIMD-based registers and pack every 32 quantization codes in a batch. In this way, it can estimate distances for 32 vectors simultaneously via a series of SIMD-based operations, which has been reported to be highly efficient by many libraries from industry~\cite{faiss, ngtlib}. For more technical details and theoretical analysis about RaBitQ and FastScan, we refer readers to their original papers~\cite{quickadc, gao2024rabitq}.

\begin{figure*}[!th]
    \centering
    \includegraphics[width=\textwidth]{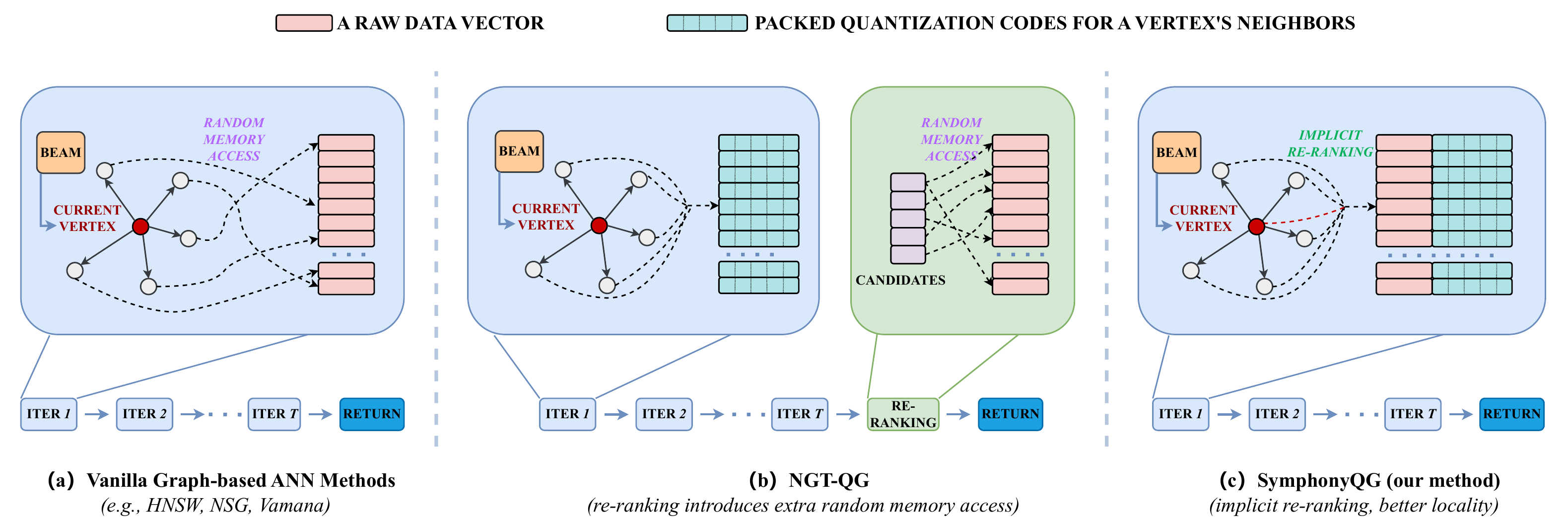}
    \vspace{-4mm}
    \caption[Caption for LOF]{Data layout and memory access pattern of vanilla graph-based methods, NGT-QG and our method.
    As illustrated, vanilla graph-based methods introduce plenty of random memory accesses when checking a vertex's neighbors in each iteration  of searching the graph since the neighbors could be stored at different places. 
    In NGT-QG,  the neighbors of each vertex are stored sequentially at the vertex's side and they can be sequentially scanned for estimating their distances with FastScan, 
    but it entails a re-ranking stage, which computes the exact distances of those candidates and incurs random memory accesses.
    Compared with NGT-QG, our method avoids the explicit re-ranking stage - instead it conducts implicit re-ranking (details will be introduced in Section~\ref{subsec: greedy search}) and thus it has no random memory accesses for re-ranking.}
    \vspace{-2mm}
    \label{fig:main}
\end{figure*}

\section{Methodology}
\label{sec: method}

In this section, we present our method SymphonyQG for ANN. 
SymphonyQG consists of a query phase and an index phase. 
During the query phase, it (1) searches on a graph-based index with FastScan and RaBitQ for boosting the search performance and (2) avoids explicit re-ranking for reducing random memory accesses. We present the query phase in Section~\ref{subsec: search}.
In particular, in Section~\ref{subsubsec: extend}, we incorporate
RaBitQ to  the scheme of searching on graph-based indices with FastScan.
Then, in Section~\ref{subsec: greedy search}, 
we design a search algorithm together with a dedicated data layout to minimize random memory accesses. 
We summarize the memory access patterns of different methods including ours in Figure~\ref{fig:main},
which shows that (1) existing vanilla graph-based ANN methods suffer from having random memory accesses within each iteration of searching the graph (Figure~\ref{fig:main}(a)); (2) the existing NGT-QG method suffers from having random memory accesses during the re-ranking stage (Figure~\ref{fig:main}(b)); and (3) our method avoids random memory accesses both within an iteration and during re-ranking (Figure~\ref{fig:main}(c)).

During the index phase, it (1) leverages FastScan for efficient graph construction and (2) builds graph indices towards better alignment with FastScan. We present the index phase in Section~\ref{subsec: indexing}.
In particular,
in Section~\ref{subsubec: efficient construction}, we propose to leverage our efficient search algorithm to accelerate the index construction
(which relies on ANN queries for finding candidates/neighbors of a vertex for creating edges). 
Moreover, in Section~\ref{subsubsec: align fastscan}, we propose to build a graph which has better alignment with FastScan. 
Specifically, we propose an adaptive pruning rule in graph construction to make sure that every vertex has its number of edges strictly being a multiple of the batch size.

We note that our method only supports ANN for Euclidean space and cosine similarity because the quantization method we use (i.e., RaBitQ~\cite{gao2024rabitq}) is designed for these two metrics.

\subsection{The Query Phase}
\label{subsec: search}

\subsubsection{Searching on Graphs with FastScan and RaBitQ for Boosting Search Performance}
\label{subsubsec: extend}
Suppose that for a given set of data vectors, a graph-based index has been built.
To boost the search performance, we adopt the scheme of searching on the graph-based index with FastScan by following NGT-QG~\cite{ngtlib}.
As is discussed in Section~\ref{sec: intro}, searching on
the graph-based indices with FastScan requires that for every vertex in the graph, a quantization code of every of its neighbors is stored on the side of the vertex (Figure~\ref{fig:batch_graph}). 
We further incorporate RaBitQ to this scheme.

We first specify the preparations needed for searching on graphs with FastScan and RaBitQ before querying.
As discussed in Section~\ref{subsec: rabitq}, RaBitQ is originally used in combination with IVF. It normalizes the raw vectors with the centroids of the clusters in IVF (i.e., using the centroids as the center vector $\mathbf{c}$) and reduces the task of estimating distances between raw vectors to one of estimating the inner product between unit vectors.
However, in graph-based indices, the centroids do not exist.
Thus, the first question is how to normalize the vectors in graph-based indices.
Recall that when searching graph-based indices, the algorithm iteratively visits a vertex and estimates distances for all its neighbors. 
A natural idea is to normalize the vectors of the neighbors with the vector of the vertex (i.e., using the vector of the vertex as the center vector $\mathbf{c}$). 
Then with the normalized data vectors, we can compute the quantization codes of RaBitQ~\footnote{It is worth noting that when a vector is a neighbor of two different vertices, its quantization codes stored on their sides are different from each other because the normalization is based on different vectors 
(e.g., in Figure~\ref{fig:batch_graph}, the red and green vertex both store quantization code of the blue vertex, but these codes are different.).}
and store them on the side of the vertex. 
Note that the normalization process can be conducted before the query phase starts since it does not rely on a query vector, and thus its cost can be shared by all queries on the dataset.
For example in Figure~\ref{fig:batch_graph}, a quantization code of the blue vertex is stored on the side of the red vertex and the quantization code is computed by normalizing the blue vector with respect to the red vector.

However, this way of normalization introduces new issues in estimating distances during querying.
In particular, recall that before efficiently estimating distances for a batch of vectors, FastScan needs to prepare LUTs (look-up tables) in prior. 
Specifically, based on  normalization, the task of estimating the distances between the raw vectors has been reduced to that of computing the inner product $\left< \mathbf{\bar x}, P^{-1}\mathbf{q}\right>$ (see Section~\ref{subsec: rabitq}) where 
$\mathbf{\bar x}$ is the bi-valued vector represented by the quantization code
and $\mathbf{q}$ is the normalized query vector with respect to a center vector $\mathbf{c}$,
i.e., $\mathbf{q}=\frac{\mathbf{q}_r - \mathbf{c}}{\|\mathbf{q}_r-\mathbf{c}\|}$. 
For the vector $\mathbf{q}$, FastScan needs to prepare LUTs such that it can compute 
$\left< \mathbf{\bar x}, P^{-1}\mathbf{q}\right>$
for a batch of quantization codes all at once.
Because $\mathbf{q}$ depends on the vector $\mathbf{c}$ for normalization, when different $\mathbf{c}$'s are used for normalization, 
different normalized query vectors $\mathbf{q}$ would be obtained, and consequently different LUTs should be prepared.
In \cite{gao2024rabitq}, RaBitQ is used in combination with IVF. 
The preparation of LUTs happens before probing a cluster and the time cost can be shared by all the vectors in the same cluster of IVF (e.g., hundreds or thousands of vectors). Thus, the amortized time cost is tiny.
However, in our circumstance, we need to prepare LUTs every time we visit a vertex (because the quantization codes 
of its neighbors are produced by using the vertex's vector as the center vector for normalization), and the cost can only be shared by 
its neighbors (e.g., 32, 64 or 128 vectors). This is very costly and could cancel out the performance gain brought by FastScan.

We solve this problem by converting the problem of computing $\left< \mathbf{\bar x}, P^{-1}\mathbf{q}\right>$ (which is dependent on a center vector $\mathbf{c}$)
to one of computing $\left< \mathbf{\bar x}, P^{-1}\mathbf{q}_r \right>$ (which is independent of the center vector $\mathbf{c}$). We have the following deductions.
\begin{align}
    \left< \mathbf{\bar x}, P^{-1} \mathbf{q}\right> &= \left< \mathbf{\bar{x}}, P^{-1}\frac{\mathbf{q}_r - \mathbf{c}}{\|\mathbf{q}_r-\mathbf{c}\|} \right> \label{eq 3.1: by definition}
    \\ &= \frac{1}{\|\mathbf{q}_r-\mathbf{c}\|}\cdot \left( \left< \mathbf{\bar{x}}, P^{-1} \mathbf{q}_r\right>  - \left< \mathbf{\bar x}, P^{-1}\mathbf{c}\right> \right) \label{eq 3.1: simplify}
\end{align}
where Equation (\ref{eq 3.1: by definition}) is due to
the definition of $\mathbf{q}$; and Equation (\ref{eq 3.1: simplify}) decomposes Equation (\ref{eq 3.1: by definition}) into two parts. \textbf{First}, $\left< \mathbf{\bar x}, P^{-1}\mathbf{c} \right>$ is independent of the query. It can be pre-computed in the index phase. 
\textbf{Second}, $\left< \mathbf{\bar x}, P^{-1}\mathbf{q}_r \right>$ is independent of the vector $\mathbf{c}$ for normalization. 
Thus, instead of preparing LUTs for computing 
$\left< \mathbf{\bar x}, P^{-1}\mathbf{q}\right>$
as the original RaBitQ does, we can prepare LUTs for computing $\left< \mathbf{\bar x}, P^{-1}\mathbf{q}_r \right>$.
As the latter is independent of $\mathbf{c}$, the LUTs can be prepared when a query comes and shared by all vectors in a database. 

In summary,
the algorithm of searching a graph-based index with FastScan and RaBitQ works as follows. 
It uses the vector of a vertex 
for normalizing its neighbors before querying.
Then, for every vertex, it computes the quantization codes of all its neighbors using RaBitQ and stores them compactly on the side of vertex.
During querying, FastScan  is used for estimating distances.
As discussed in an existing study~\cite{ciaccia2002searching}, 
the effectiveness of using an estimated to prune out objects relies on the similarity 
between the distribution of estimated distance and the distribution of the "true" distance.
Our method uses estimated distances produced by RaBitQ to guide the search process during querying.
Because RaBitQ provides an unbiased estimator with a theoretical error bound~\cite{gao2024rabitq}, 
the estimated distance used in our method has a similar distribution with the "true" distance.
Therefore, it is effective for our method to prune out objects to improve the query efficiency,
which is aligned with our experimental results.

\subsubsection{ Avoiding Explicit Re-Ranking for Reducing Random Memory Accesses}
\label{subsec: greedy search}
Recall that during the process of searching on graph-based indices, vanilla greedy beam search computes exact distances based on raw vectors, which requires random memory accesses within each iteration of searching
(see Section~\ref{subsec:ann and graph} and Figure~\ref{fig:main}(a)).
When searching on graph-based indices with FastScan, the algorithm estimates distances of the neighbors of a vertex based on the quantization codes stored at its side, which 
avoids random memory accesses within each iteration (see Figure~\ref{fig:main}(b)).
However, because the searching is based on the estimated distances, this entails a re-ranking step to achieve reasonable recall.
Specifically, in NGT-QG, at the end,
it accesses the vectors which have the smallest estimated distances to the query,
computes their exact distances and finds the nearest neighbor accordingly.
Because for computing exact distances, it needs to access raw vectors, this introduces random memory accesses.
In order to eliminate 
these random accesses,
we propose to conduct re-ranking implicitly during searching on graph-based indices, which we explain as follows.

Recall that as discussed in Section~\ref{subsubsec: extend}, our method 
uses the vector of a vertex for normalizing the vectors of all its neighbors. 
In order to estimate distances for all its neighbors based on Equation (\ref{eq: normalization}), this requires to compute the exact distance between the vector of the vertex and the query vector 
(i.e., $\|\mathbf{q}_r-\mathbf{c}\|$ in Equation~(\ref{eq: normalization}), where $\mathbf{c}$ corresponds to the vector of the vertex).
Therefore, we re-use this exact distance to maintain the NN that has been searched so far while using the estimated distances for searching graph-based indices. With a specialized data layout (which will be specified latter and illustrated in  Figure~\ref{fig:main}(c)), this does not introduce extra random memory accesses.
When the search process terminates, the vector with smallest exact distance maintained in the algorithm is returned as the result of the ANN query without an extra re-ranking step. 

The above idea would help remove the random accesses of re-ranking, but it introduces accuracy loss. 
Specifically, in the above process, we update the NN only when we visit a vector and compute its exact distance. 
Note that a vector would be visited only when it has the smallest estimated distances during searching. 
If the distance of a true NN is over-estimated, the NN is likely to be missed. 
To resolve this issue, we propose to conduct greedy beam search using \emph{multiple} estimated distances for every vector.
Specifically, during the conventional greedy beam search with FastScan, in each iteration, the algorithm visits a vertex and estimates
the distance for each of its neighbors. 
Each neighbor along with its estimated distance is appended into the beam set if the neighbor has not been in the set. 
In contrast, in our method, we always append the neighbor along with its estimated distance into the beam set 
unless it has been visited.
As a result, the same vertex may be added to the beam set multiple times (along with different estimated distances).
The rationale is as follows.
 
Recall that based on the theoretical guarantee of RaBitQ, 
every estimated distance is unbiased and error-bounded. 
Thus, when multiple estimated distances are used for a vector during searching, the search algorithm will have the following properties.
(1) For a true NN, in each estimation, RaBitQ has 50\% probability to produce an under-estimation of its distance. 
When multiple estimations are used, it is highly likely that there is at least one under-estimation.
Note that the true NN has the smallest exact distance among all vectors. Therefore, an under-estimation of its distances is highly likely to be the smallest in the beam set in a certain iteration. This leads to a visit to the true NN and thus, the true NN is not missed. 
(2) For other vectors, RaBitQ guarantees that the estimated distances are error-bounded. When multiple estimations are used, it is guaranteed that all the estimated distances are within an error bound, i.e., 
all of the multiple estimated distances is unlikely to deviate significantly from the true distances due to the bound. Therefore, when the true distance of a vector is larger than that of the NN by a clear margin, it is less likely that it would be visited due to a severe under-estimation.
On the other hand, when the true distance of a vector is slightly larger than that of the NN, the vector is very close to the query. Visiting the vector could help to find a true NN on the graph. 
Thus, based on this strategy, the accuracy is secured.
The ablation study in  Section~\ref{subsec: ablation} verifies the effectiveness of this technique.

\begin{algorithm}
\caption{ Querying} \label{alg:hybrid_search}
\begin{algorithmic}[1]

\Require{A query vector, the beam size $n_b$, a graph-based index $\mathbf{G}$, an entry point $e$}
\Ensure{The nearest neighbor of the query}
\State Initialize the beam set $\mathcal{S}$ and $NN$ with the entry point $e$
\While{$\exists$ an unvisited vertex in $\mathcal{S}$}
    \State Find the unvisited vertex $p$ with the smallest estimated distance from $\mathcal{S}$ and mark $p$ as visited
    \State Compute the exact distance of $p$ and update $NN$
    \State Estimate distances for the neighbors of $p$ in $\mathbf{G}$ via FastScan
    \State Append all 
    unvisited neighbors
    along with their estimated distances into $\mathcal{S}$ and cut off the size of $\mathcal{S}$ to $n_b$
\EndWhile

\State \Return{$NN$}

\end{algorithmic}
\end{algorithm}

\subsubsection{Summary of Querying Algorithm}
\label{subsec: summary}
We summarize the querying algorithm in Algorithm~\ref{alg:hybrid_search}. 
Given a query vector, a hyperparameter beam size $n_b$, a graph-based index $\mathbf{G}$ and an entry point $e$, the algorithm starts with initializing the beam set $\mathcal{S}$ and the nearest neighbor $NN$ with the entry point (line 1). 
It then proceeds querying in iterations (line 2-6). In each iteration, it first finds the unvisited vertex with the smallest approximate distance $p$ and computes its exact distance (line 3-4).
Next, it estimates distances for its neighbors with FastScan and appends all the unvisited neighbors along with the estimated distances to the beam set (line 5-6).
The algorithm terminates when all the vertices in $\mathcal{S}$ have been visited and returns the nearest neighbor. 

Based on the querying algorithm above, we propose a dedicated data layout in order to minimize the random memory accesses, which is presented as follows.
Overall, the index consists of $n$ data objects.
Each data object is assigned with a sequential block of memory which stores (1) its raw data vector, 
(2) the data for FastScan  (including the quantization codes and the pre-computed factors needed by RaBitQ, e.g., $\| \mathbf{o-c}\|$) 
and (3) the neighbors of the graph (see Figure~\ref{fig:main}(c)
~\footnote{Here, we omit the data of neighbors in Figure~\ref{fig:main} to provide a clearer comparison among different methods.
In practice, the neighbors of each vertex are stored right after the data for FastScan in our method for better cache locality.}).
During querying, when visiting a certain vertex on the graph in an iteration, 
the algorithm accesses the corresponding block memory of the vertex.
Clearly, the algorithm accesses a continuous segment in memory when visiting a vertex in each iteration, which leads to good cache locality.
Based on this data layout, we summarize and compare the memory access pattern of the vanilla graph-based methods, NGT-QG and our method in Figure~\ref{fig:main}.
We would like to highlight that compared with NGT-QG, our method eliminates the re-ranking at the end of the algorithm. 
This avoids the costly random memory accesses.

\subsubsection{Implementation Techniques}
\label{subsec: details}
Besides the aforementioned techniques in terms of algorithms, 
we would also like to mention the following techniques in our implementation.

\vskip5pt\noindent\textbf{Memory Prefectching.}
Although the random memory accesses within an iteration have almost been eliminated, there could still be cache miss between iterations, i.e., the sequential piece of data for the vertex to visit in the next iteration might not be cached.
In order to further relieve the issue,
we prefetch the data of the vector which has the smallest distance 
when updating the beam set (line 6 in Algorithm~\ref{alg:hybrid_search}).

\vskip5pt\noindent\textbf{More SIMD Acceleration.}
In \cite{gao2024rabitq}, RaBitQ only uses SIMD (i.e., FastScan) for computing 
$\left<\mathbf{\bar x}, P^{-1}\mathbf{q} \right>$
and supports only till AVX2.
We conduct more extensive optimizations with SIMD in the whole process of the algorithm, e.g., using SIMD to accelerate the computation of Equation (\ref{eq: normalization}), and provide the support of AVX512.

\vskip5pt\noindent\textbf{Fast Johnson-Lindenstrauss Transformation.}
For efficiently implementing the random rotation (Johnson-Lindenstrauss Transformation~\cite{johnson1984extensions}, JLT) for RaBitQ, 
we apply an algorithm of Fast JLT based on Fast Hadamard Transformation~\cite{falconn, ffhtlib, SRHT}. 
This improves the time complexity of random rotation from $O(D^2)$ to $O(D\log D)$, where $D$ denotes the number of dimensions of a vector.

\subsection{The Index Phase}
\label{subsec: indexing}

\subsubsection{Leveraging FastScan for Building Graph-based Indices Efficiently}
\label{subsubec: efficient construction}
In many graph-based ANN methods~\cite{hnsw,nsw,nsg,2019diskann,tau-mng,onng,fu2016efanna,wang2023graph}, 
constructing edges for each vertex can be roughly divided into two stages.
In stage one, the algorithm finds several vertices as the candidates for a vertex's neighbors.
In stage two, 
a pruning rule is applied on the candidates. 
Finally, an edge between each candidate that has not be pruned and the vertex will be created.
For example, NSG~\cite{nsg} constructs its graph in the scheme of initializing a graph and adjusting it.
Specifically, 
for every vector, it conducts ANN query on the initialized graph.
Then, the results of ANN query are taken as the candidates.
A new graph is then built after applying a pruning rule on the candidates.

We notice that our efficient search algorithm can be easily applied to accelerate the above scheme of graph construction.
Specifically, the above construction algorithm conducts ANN query for every vector to generate candidates, 
during which, our efficient search algorithm can be applied to accelerate the ANN query.
The detailed process of our index construction is presented as follows. We consider randomly initializing a graph and adjusting it iteratively.
In each iteration, for the graph produced in the last iteration, 
we first prepare the data for FastScan (e.g., quantization codes).
Then, we apply our search algorithm for finding candidates for every vertex.
For each vertex, we use the pruning rule of NSG~\cite{nsg} to select at most $R$ vertices as the vertex's new neighbors (the neighbors in the graph of the last iteration will not be retained in the new graph).
After all vertices have their new neighbors, we adjust edges in the graph and start the next iteration.
Note that the graph index is adjusted at the end of each iteration and is unchanged within an iteration. 
This indicates that the candidate generation and pruning of different vertices are independent to each other. 
Thus, the indexing algorithm in each iteration is highly parallelable.
According to our experimental results in Section~\ref{subsec: ablation}, compared with the original algorithm of constructing NSG, 
the above indexing algorithm brings at least 2.5x speed-up in indexing without harming the search performance of the constructed graphs. It is also worth emphasizing that our method is at least 8x faster than NGT-QG in indexing.

\begin{figure}[htbp]
    \centering
    \includegraphics[width=\linewidth]{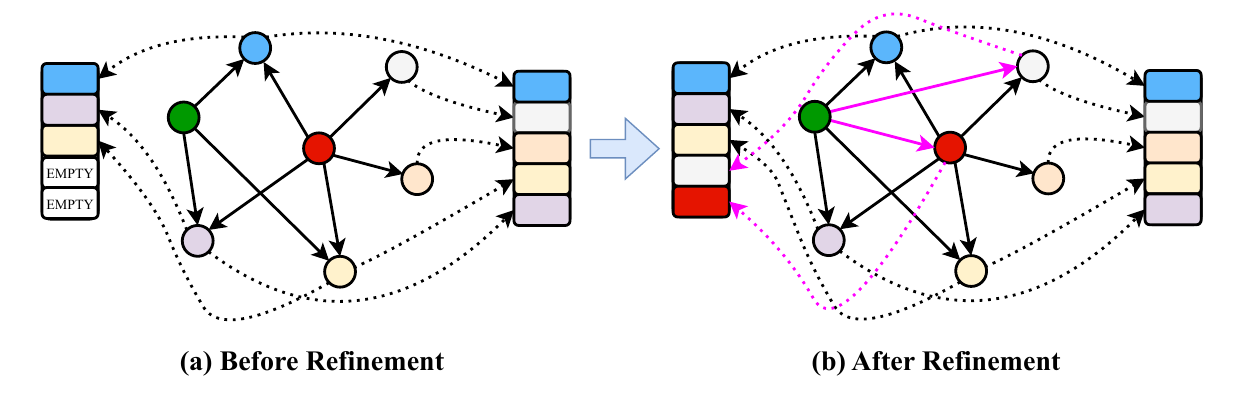}
    \vspace{-4mm}
    \caption{
    Refining a graph towards better alignment with FastScan by supplementing edges.
    Here, we assume $R$ and the batch size are 5 for the simplicity of illustration.
    The out-degree of the green vertex is only 3 in the original graph because the original pruning rule is too restrictive.
    After supplementing edges with an adaptive pruning rule, its out-degree is guaranteed to be 5.
    For the red vertex, its out-degree is already aligned with the batch size, so we do not need to supplement edges for this vertex.
    }
    \Description[<short description>]{}
    \label{fig:edge supp}
    \vspace{-2mm}
\end{figure}

\subsubsection{Refining Graph-based indices towards Better Alignment with FastScan}
\label{subsubsec: align fastscan}
Based on the above algorithm, we have built a graph based on an existing pruning rule of NSG.
However, as discussed in Section~\ref{sec: intro}, 
FastScan requires to estimate distances in a batch for 32 vectors simultaneously.
Without a specialized design, it is likely that the out-degree of a vertex in the constructed graph is not a multiple of the batch size and thus, leads to a waste of computation.
We note that the graph cannot have the desired number of edges because the pruning rule is sometimes too restrictive, i.e., after pruning, the number of edges has already been smaller than a target which we denote by $R$.

According to the experimental results in Section~\ref{subsec: ablation}, based on the pruning rule of NSG, the average out-degree of the graph is less than $R$ on all tested datasets (e.g., the average out-degree of graph on the MSong dataset is 19.8 when $R=32$).
To solve this issue, we propose to further refine the graph by supplementing edges in the graph 
such that the number of edges of each vertex is strictly aligned with the batch size(see an example in Figure~\ref{fig:edge supp}).

Specifically, we first build edges based on the original pruning rule of NSG~\cite{nsg}. For a vertex, if the number of its edges is smaller than $R$, we re-consider the candidates that have been pruned and adaptively relax the pruning rule on them such that every vertex has exactly $R$ edges.
In particular, recall that according to the analysis in NSG~\cite{nsg}, its pruning rule ensures the diversity of the edges of a vertex, i.e., the angle between any two out-edges of a vertex in the high-dimensional space is at least $60^{\circ}$.
Given the pruned candidates, we apply an \emph{angle-based} pruning rule, i.e., for a candidate of the vertex, 
if there exists another candidate 
whose distance to the vertex is smaller and the angle between the edges of the candidates is smaller than a given threshold, 
then we prune the candidate. 
It is clear that the larger the threshold is, it is more likely that a candidate is pruned and subsequently, the smaller the candidate set is after pruning.
As the size of the candidate set after pruning is \emph{monotonic} with respect to the threshold,
we apply \emph{binary search} on the threshold. 
It adaptively finds the most restrictive pruning rule (i.e., the largest threshold for pruning) such that a vertex can have exactly $R$ neighbors. 
Note that the pruning rule is adaptive in the sense that different vertices may have different thresholds for pruning. 
Thus, it is ensured that every vertex in the graph strictly has a preset number of out-edges
~\footnote{In the extreme case that the number of candidates is smaller than the maximum out-degree, our method will use randomly selected neighbors to make sure every vertex's out-degree is a multiple of 32.}.
This further ensures that the degree of every vertex is aligned with FastScan.

Note that because in FastScan, the distances are estimated batch by batch,
supplementing edges for a graph to fill a batch does not increase the time cost of distance estimations. On the other hand, the more edges a graph has, the less likely the true NN is missed during graph-based searching. Thus, the algorithm achieves better time-accuracy trade-off. 
According to the ablation study in Section~\ref{subsec: ablation},our graph refinement
strategy brings consistent improvement in search performance on all the tested datasets.

\begin{algorithm}
\caption{Indexing} \label{alg:indexing}
\begin{algorithmic}[1]
\algloopdefx{Parallel}[1]{\textbf{for} #1}

\Require{A dataset $\mathcal{D}$, the out-degree $R$, the \# of iterations $t$}
\Ensure{A graph-based index $\mathbf{G}$ with the data for FastScan}
\State Randomly initialize a graph $\mathbf{G}$ 

\For{$1 \le i \le t$}
    \For{all vertices}
        \State Prepare data for FastScan based on $\mathbf{G}$
    \EndFor
    \For{all vertices}
        \State Search on $\mathbf{G}$ to find candidates
        \State Prune candidates to obtain new neighbors
    \EndFor
    \State Adjust $\mathbf{G}$
\EndFor
\State Supplement edges for vertices whose out-degree $< R$
\State Update the data for FastScan based on $\mathbf{G}$
\State \Return{$\mathbf{G}$ with the data for FastScan}

\end{algorithmic}
\end{algorithm}

We summarize the indexing of our method in Algorithm~\ref{alg:indexing}.
The graph is randomly initialized to a graph, where every vertex has $R$ neighbors.
Then the graph is adjusted iteratively (line 2-8).
In each iteration, the algorithm prepares the data for FastScan based on $\mathbf{G}$ (line 3-4).
Then it uses our querying algorithm to find the candidates of neighbors for all vertices and applies NSG's pruning rule on these candidates
(line 5-7).
At the end of each iteration, it adjusts the graph with new neighbors for each vertex (line 8).
As discussed, the candidate generation and pruning are highly parallelable across different vertices. 
After $t$ iterations ($t=3$ or 4 in practice),
it finds vertices whose out-degree is less than $R$ and supplement edges for them (line 9).
At the end of indexing, 
it updates the data used for FastScan based on the final graph and returns the index (line 10-11).

\subsection{The Summary}
\label{subsec: method summary}
In summary, SymphonyQG handles ANN query by using RaBitQ and FastScan to search on graph-based indices.
During querying, we incorporate RaBitQ to graph-based indices by resolving issues in normalization and efficient distance estimation.
To reduce the cost of random memory access, we design the implicit re-ranking strategy and eliminate the  
explicit re-ranking step of NGT-QG. 
To handle the issue of possible accuracy loss caused by implicit re-ranking, we propose a search strategy based on multiple estimated distances.
As for indexing, we use our efficient search algorithm to accelerate the candidate generation process of index construction.
Moreover, considering the in-batch computation of FastScan,
we design a graph refinement strategy  with an adaptive pruning rule, which makes sure the out-degree of every vertex is aligned with the batch size.
Experimental studies in Section~\ref{sec: experiments} shows that based on all the proposed techniques, 
SymphonyQG establishes the new state-of-the-art in terms of the time-accuracy trade-off
on all tested real-world datasets.
For example, at 95\% recall ($K=10$),
our method achieves 1.5x-4.5x QPS compared with the most competitive baselines (which differ from dataset to dataset) and achieves 3.5x-17x compared with the classical library HNSWlib on all tested datasets. 
At the same time, the indexing
of our method is at least 2.5x faster than NSG and is at least 8x faster than NGT-QG.

\smallskip\noindent\textbf{Overhead of Memory Consumption.}
\label{subsubsec: space analy}
Despite the promising performance of SymphonyQG in terms of the accuracy-efficiency trade-off, we would like to note that the method brings some overhead of memory consumption, whose details are presented as follows.
Let $D$ and $R$ be the dimensionality and the out-degree of every vector respectively.
First of all, like vanilla graph-based methods, for every vector, SymphonyQG uses $32\times D$ bits ($D$ floating-point numbers) to store the raw vectors and $32\times R$ bits ($R$ integers) to store a vector's neighbors.
In addition, recall that by following NGT-QG~\cite{ngtqgblog}, for every vector, SymphonyQG creates quantization codes for all its $R$ neighbors. Each quantization code takes $D$ bits~\cite{gao2024rabitq}, and thus, the extra space
for storing the codes is $D\times R$ bits. 
Therefore, the total space cost of our index is $n \times (32D + 32R + DR)$ bits, where $n$ is the number of vertices.
Under the typical settings of the out-degree ($R=32$ and $64$), the overhead for storing the codes is 1x and 2x of that for storing the raw  vectors respectively. 

\section{EXPERIMENTS}
\label{sec: experiments}

\begin{figure*}[h]
    \centering
    \includegraphics[width=\textwidth]{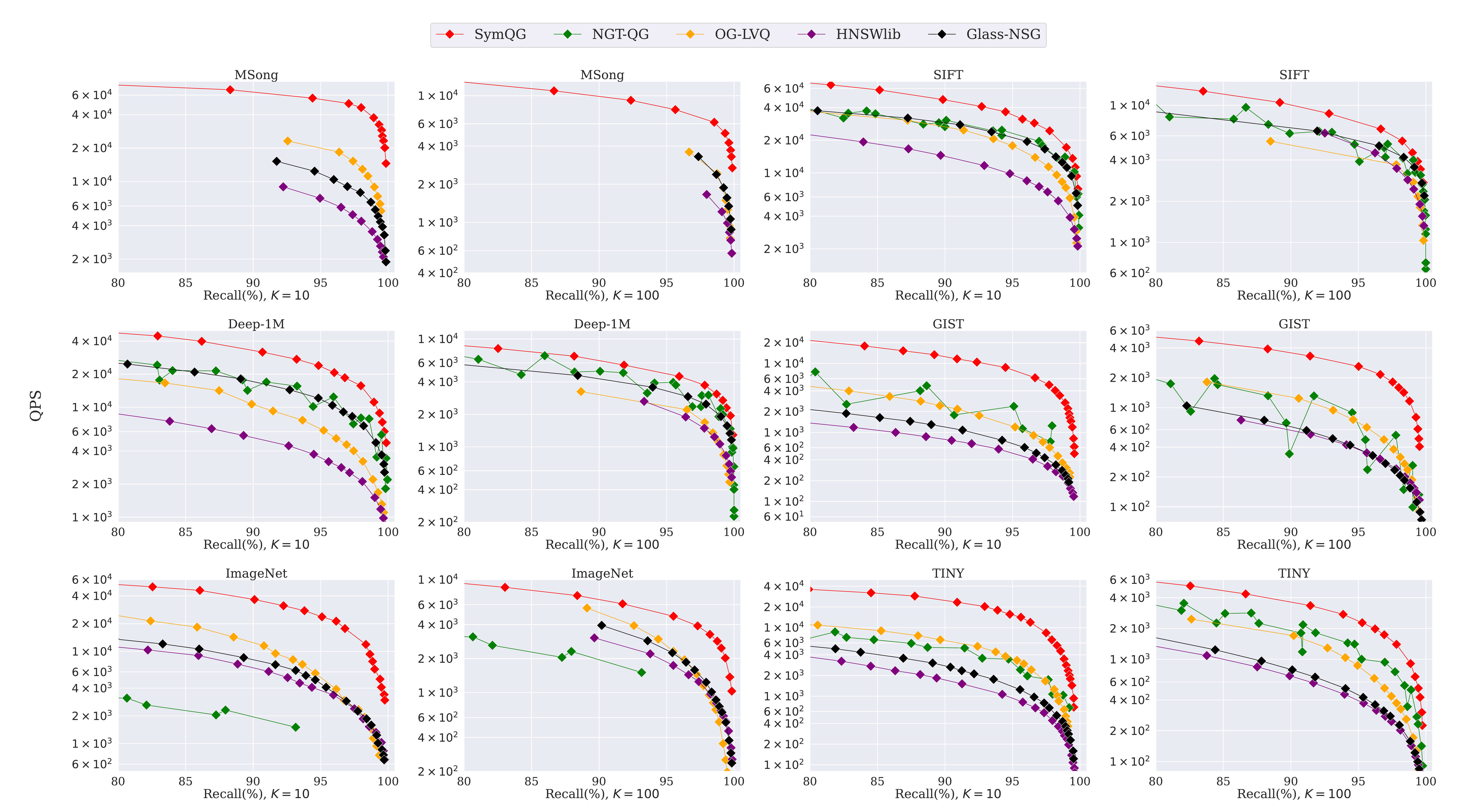}
    \vspace{-4mm}
    \caption{Comparison of all methods on 6 real-world datasets (QPS-Recall). 
    The missing curve of a certain method indicates that it fails to achieve at least 80\% recall on the dataset.}
    \vspace{-2mm}
    \label{fig:main exp}
\end{figure*}

\begin{figure*}[h]
    \centering
    \includegraphics[width=\textwidth]{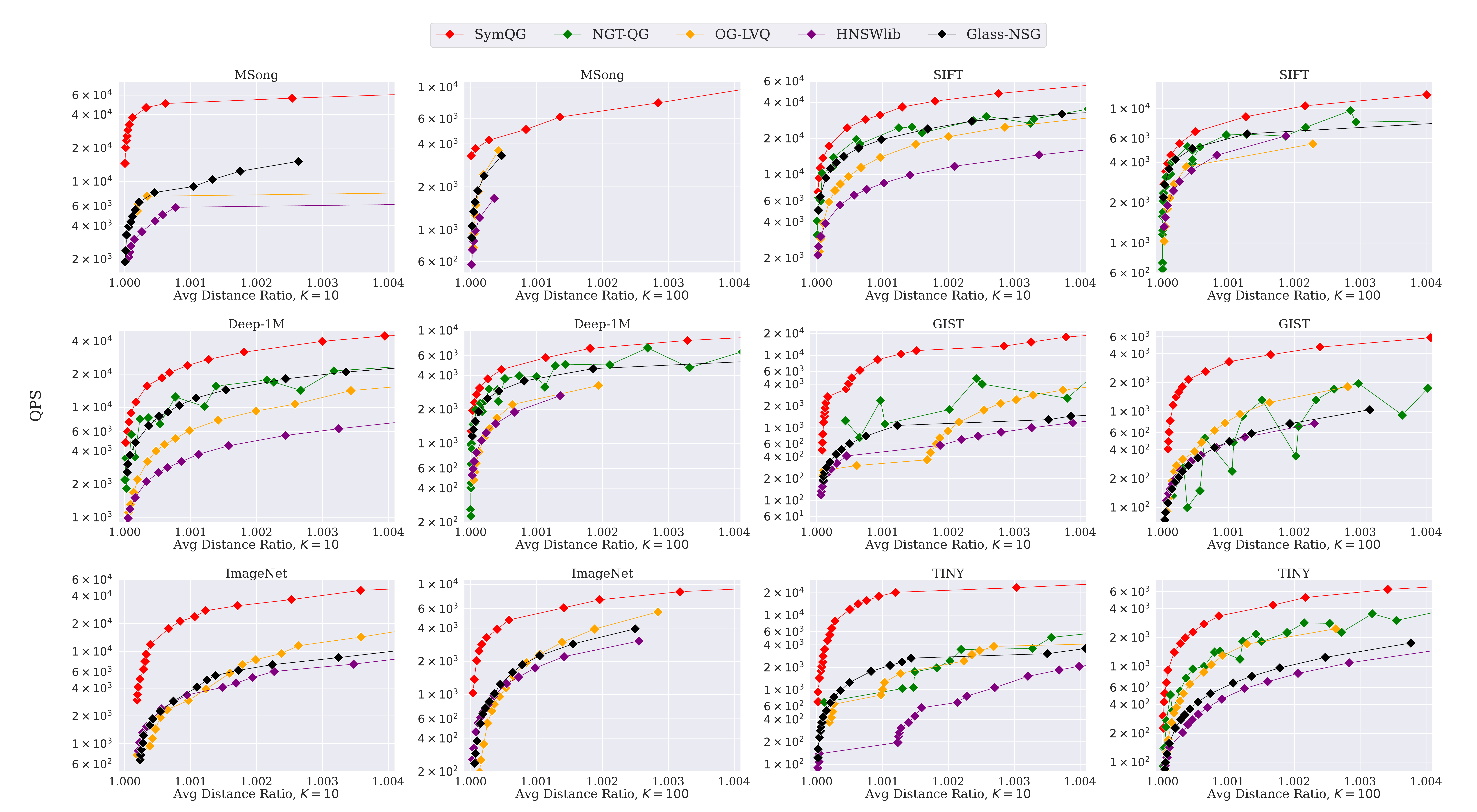}
    \vspace{-4mm}
    \caption{Comparison of all methods on 6 real-world datasets (QPS-Ave Distance Ratio).
    The missing curve of a certain method indicates that it fails to achieve at least 1.004 average distance ratio on the dataset.}
    \vspace{-2mm}
    \label{fig:ratio}
\end{figure*}

We provide the experimental setup in Section~\ref{subsec: exp setup} and the experimental results which involve four folds in Section~\ref{subsec: exp results}.
(1) In Section~\ref{subsec: exp moderate scale}, we evaluate the search performance of our method by comparing it with the state-of-the-art open-source libraries on real-world datasets. 
We also report the indexing time and memory footprint of every method.
(2) In Section~\ref{subsec: scalability}, we test the scalability of our method on a dataset with 100 million data vectors.
(3) In Section~\ref{subsec: ablation}, we verify the effectiveness of the proposed techniques via an ablation study.
(4) In Section~\ref{subsubsec: parameter study}, we conduct the parameter study for our newly introduced parameter.

\subsection{Experimental Setup}
\label{subsec: exp setup}

\smallskip\noindent\textbf{Datasets.}
To evaluate the search performance of different methods, 
we adopt 6 moderate-scale (million-scale) datasets including
MSong,
SIFT,
Deep-1M,
GIST,
ImageNet
and TINY.
We also test the scalability of our method on two large-scale datasets
Deep-100M and MSTuring-100M~\footnote{We  adopted the first 10K query vectors from the query set of MSTuring-100M.}.
These datasets cover diverse dimensionality, sizes and similarity metrics,
and they have been widely used in existing studies~\cite{hnsw, nsg, intellvq, dobson2023scaling_billion_benchmark, annbenchmark, nnsbench, bigann2021, bigann2023} to benchmark ANN algorithms and libraries.
All these datasets provide both the data vectors and a set of query vectors for evaluation. 
The detailed statistics of the datasets are presented in Table~\ref{tab: datasets}.

\smallskip\noindent\textbf{Machine Configuration}.
We run all experiments on a server with two Intel(R) Xeon(R) Gold 6418H@4.0GHz CPUs and 1TB DDR5 memory (@4800MT/s) under Ubuntu 22.04 LTS with GCC 11.4.0 compiler.
For all methods,  by following most of the existing studies and benchmarks~\cite{graph_benchmark, nsg, 2019diskann, tau-mng, hnsw, dobson2023scaling_billion_benchmark, annbenchmark}, 
the search performance is evaluated in a single thread and the indexing time is measured using multiple threads (96 threads, 48 cores with our server).

\smallskip\noindent\textbf{Metrics of Query Performance.}
Following existing studies and benchmarks ~\cite{annbenchmark, hnsw, nsg, 2019diskann, dobson2023scaling_billion_benchmark, marco_survey, marco_SISAP}, we measure the query efficiency by the number of queries responded per second (QPS), 
and we measure the query accuracy by recall and average distance ratio.
The recall is defined as $\frac{|G \cap S|}{K}$, 
where $G$ denotes groundtruth of $K$ nearest neighbors, and $S$ denotes search results of an ANN algorithm.
The average distance ratio (ADR) is the average of the distance ratios of the ANN query results wrt the groundtruth of $K$ nearest neighbors.
To show the trade-off between accuracy and efficiency of different methods, 
we draw QPS-recall curve and QPS-ADR curves.

\begin{table}[!htbp]
    \centering
    \small
    \vspace{-2mm}
    \caption{Statistics of the datasets. $N$ and $N_q$ denote the number of data vectors and the number of queries, respectively. 
    $D$ denotes the dimensionality of vectors}.
    \vspace{-2mm}
    \label{tab: datasets}
    \setlength{\tabcolsep}{0pt}
    \begin{tabularx}{\columnwidth}{l *{5}{>{\centering\arraybackslash}X}}
    \toprule
        Dataset & $N$ & $N_q$ & $D$ & Size (GiB) & Distance \\
    \hline
        MSong & $\sim$1M & 200 & 420 & 1.6 & Euclidean \\
        SIFT & 1M & 10,000 & 128 & 0.5 & Euclidean \\
        Deep-1M & 1M & 1,000 & 256 & 0.9 & Cosine \\
        GIST & 1M & 1,000 & 960 & 3.6 & Euclidean \\
        ImageNet & $\sim$2.3M & 200 & 150 & 1.4 & Euclidean \\
        TINY & 5M & 200 & 384 & 7.2 & Euclidean \\
        Deep-100M & 100M & 10,000 & 96 & 37.8 & Cosine \\
        MSTuring-100M & 100M & 10,000 & 100 & 40.0 & Cosine \\
    \bottomrule
    \end{tabularx}
    \vspace{-2mm}
\end{table}

\smallskip\noindent\textbf{Methods and Parameters.}
We compare our method with four well-optimized graph-based open-source libraries including NGT-QG (Yahoo Japan)~\cite{ngtlib},
OG-LVQ (Intel)~\cite{intellvq}, Glass-NSG (Zilliz)~\cite{glass} and HNSWlib~\cite{hnsw,hnswlib}.
For all methods, we conduct parameter search to find the optimal setting of the parameters for 
indexing~\footnote{We build multiple indices and test their QPS-recall trade-off. 
Based on the results, we adopt the parameter for indexing which achieves the highest QPS at 95\% recall.}.
Then, we fix the optimal parameters for indexing and plot the QPS-recall curve by varying the parameters for querying.
The details of the optimal parameter settings for each dataset are provided in 
the appendix 
due to the limitation of space.
\underline{(1)~SymphonyQG} (SymQG for short) is our method. 
There are three parameters during indexing of SymphonyQG,
the maximum out-degree of the graph $R$, a parameter that controls the number of candidates during graph construction named $EF$ and the number of iteration for graph construction $t$ (more details can be found in Section~\ref{subsec: indexing}).
During querying, the beam size $n_b$ is varied to control the time-accuracy trade-off. 
We fix $EF$ to 400 and vary $R$ among 32, 64, 128.
We set $t$ as 4 for MSTuring-100M and 3 for other datasets.
\underline{(2)~NGT-QG}~\cite{ngtlib} is a method published in the open-source library NGT by Yahoo Japan. 
It is the originator of the QG framework, which pioneers in searching on graph-based indices with FastScan and has been reported to be one of the state-of-the-art 
libraries in the well-acknowledged ANN-Benchmark~\cite{annbenchmark}.
The indexing and querying of NGT-QG involves many parameters. 
We refer readers to the library~\cite{ngtlib} for detailed information.
We conduct parameter search among its suggested parameter settings in ANN-Benchmark~\cite{annbenchmark}.
\underline{(3)~OG-LVQ}~\cite{intellvq} is a method published in the open-source library SVS by Intel.
It is based on graph-based indices and scalar quantization.
The indexing parameters include a) the parameters for graph-based indices including $R$ and $EF$ like our method anda parameter 
$\alpha$ which controls the pruning of its graph-based index Vamana~\cite{2019diskann} and b) the number of bits for scalar quantization.
We follow \cite{intellvq} by fixing $\alpha=1.2$ and decide $R$ and $EF$ in the same way of our method.
As for the number of bits for scalar quantization, we search parameters among its provided settings including LVQ-8, LVQ-4x8, and LVQ-8x8.
During querying, similar to our method, the beam size $n_b$ is varied 
to control the time-accuracy trade-off.
\underline{(4)~Glass}~\cite{glass} is an open-source graph-based ANN library published by Zilliz.
It has been reported to be one of the state-of-the-arts in ANN-Benchmarks~\cite{annbenchmark}.
In our experiment, 
we use NSG~\cite{nsg} provided in Glass~\footnote{We fix several bugs for this library so that it can run normally 
on the large-scale dataset Deep-100M.}.
The indexing parameters include $R$ and $EF$ like our method. 
Besides, it involves a parameter which indicates the optimization level. We observe that a high optimization level may cause disastrous failure. Thus, for each dataset, we evaluate the search performance based on all optimization levels and report the best results. 
$R$ and $EF$ are decided in the same way of our method. The time-accuracy trade-off in the query phase is controlled by the beam size $n_b$.
\underline{(5)~HNSWlib}~\cite{hnswlib} 
offers an efficient implementation of the classical graph-based ANN method HNSW~\cite{hnsw}.
The indexing parameters of HNSW include $EF$ and maximum out-degree $M$ similar to our method, and they are decided in the same way of our method. 
The time-accuracy trade-off in the query phase is controlled by the beam size $n_b$.
As all the libraries provide convenient Python interfaces, we conduct all experiments based on the Python bindings.
We note that in recent years, there are also many scientific studies which explore the opportunities of optimizing different aspects of ANN algorithms~\cite{adsampling, zhao2023towards,nssg,chen2023finger,tau-mng}. 
However, these scientific studies usually focus on 
a certain component in an algorithm while spending less effort on engineering optimizations on the whole library. 
This makes their empirical performance less competitive when compared with well-optimized libraries such as  NGT~\cite{ngtlib} (Yahoo Japan), SVS~\cite{intellvq} (Intel) and Glass~\cite{glass} (Zilliz), 
which are mostly from industry and have been included for comparison in this paper. 
We, thus, exclude these studies from comparison.

\begin{table}[!htbp]
    \centering
    \caption{Indexing time (minutes). 
    The results of NGT-QG are not shown on two large-scale datasets since it fails to build an index with 1TB memory within 1 day.
    }
    \label{tab: indexing time}
    \vspace{-4mm}
    \small
    \setlength{\tabcolsep}{0pt}
    \begin{tabularx}{\columnwidth}{l *{5}{>{\centering\arraybackslash}X}}
    \toprule
        & SymQG & NGT-QG & OG-LVQ & Glass-NSG & HNSWlib \\
    \hline
        MSong       & 0.8 & 97.3 & 2.2 & 2.1 & 1.0 \\
        SIFT        & 0.4 & 6.5 & 1.1 & 1.3 & 0.4 \\
        Deep-1M & 0.8 & 8.1 & 2.3 & 2.1 & 0.8 \\
        GIST        & 2.6 & 27.3 & 6.5 & 6.1 & 3.3 \\
        ImageNet    & 1.6 & 14.3 & 3.3 & 3.4 & 1.2 \\
        TINY        & 7.6 & 142.2 & 19.9 & 19.5 & 9.7 \\
        Deep-100M & 51.2 & - & 182.5 & 290.6 & 47.6 \\
        MSTuring-100M & 92.6 & - & 165.8 & 304.2 & 59.8 \\
    \bottomrule
    \end{tabularx}
    \vspace{-2mm}
\end{table}

\begin{table}[!htbp]
    \centering
    \caption{Memory footprint (GiB). 
    The results of NGT-QG are not shown on two large-scale datasets since it fails to build an index with 1TB memory within 1 day.
    }
    \label{tab: indexing size}
    \vspace{-4mm}
    \small
    \setlength{\tabcolsep}{0pt}
    \begin{tabularx}{\columnwidth}{l *{5}{>{\centering\arraybackslash}X}}
    \toprule
        & SymQG & NGT-QG & OG-LVQ & Glass-NSG & HNSWlib \\
    \hline
        MSong       & 4.0 & 12.0 & 1.0 & 1.7 & 1.9 \\
        SIFT        & 1.5 & 5.2 & 0.5 & 0.7 & 0.7 \\
        Deep-1M & 3.9 & 10.8 & 0.6 & 1.2 & 1.3 \\
        GIST        & 12.5 & 22.9 & 1.7 & 3.8 & 4.2 \\
        ImageNet    & 8.1  & 16.0 & 1.2 & 1.9 & 2.1 \\ 
        TINY        & 32.1 & 84.2 & 4.1 & 8.4 & 8.9 \\
        Deep-100M & 140.3 & - & 30.9 & 51.6 & 59.2 \\
        MSTuring-100M & 140.3 & - & 33.5 & 53.2 & 59.4 \\
    \bottomrule
    \end{tabularx}
    \vspace{-2mm}
\end{table}

\subsection{Experimental Results}
\label{subsec: exp results}
\subsubsection{Performance Study}
\label{subsec: exp moderate scale}
In this section, we evaluate the search performance as well as the indexing time and memory footprint of each method on real-world datasets.

\smallskip\noindent\textbf{Search Performance.}
We first evaluate the search performance for all methods on moderate-scale datasets.
As discussed, we build indices with the optimal parameter setting for each dataset and vary parameters in the query phase 
(e.g., the beam size $n_b$) to plot the QPS-recall curve and QPS-ADR curve.
In particular, the search performance is evaluated under two different settings including $K=10$ and $K=100$.
Figure~\ref{fig:main exp} plots QPS-recall curves (upper-right is better), 
and Figure~\ref{fig:ratio} plots QPS-ADR curves (upper-left is better) on six moderate-scale datasets.
We have the following key observations.
(1) In general, the search performance of SymQG surpasses all baseline methods.
Compared to other methods, SymQG achieves a better QPS-recall trade-off and better QPS-ADR trade-off on all tested datasets.
When $K=10$,
our method surpasses the most competitive baselines (which differ from dataset to dataset) by 1.5x-4.5x in QPS at 95\% recall on all the tested datasets.
Moreover, SymQG outperforms the classical baseline HNSWlib by 3.5x-17x in QPS at 95\% recall on all tested datasets.
When $K=100$, our method still shows the best search performance on tested datasets.
As for QPS-ADR curves presented in Figure~\ref{fig:ratio}, 
our method achieves a higher QPS than all baselines at the same ratio for both $K=10$ and $K=100$ for all tested datasets, 
which is consistent with the QPS-recall curves.
(2) As discussed in Section~\ref{sec: intro},
our method can handle ANN query of datasets on which NGT-QG fails.
As illustrated in Figure~\ref{fig:main exp}, we observe NGT-QG cannot achieve at least 80\% recall on the MSong dataset.
This can be explained by the disastrous failure of PQ on the dataset as has been reported in \cite{gao2024rabitq}.
It is worth noting that on ImageNet, NGT-QG also incurs severe performance drop.
In contrast, our method has more robust performance across different datasets.
(3) The curves of NGT-QG are not as smooth as those of other methods. This is because during searching,
it involves two parameters, one for searching and the other for re-ranking, and it needs to tune these parameters simultaneously.
This introduces new difficulties of parameter tuning when using NGT-QG in practice.
We refer readers NGT-QG's library~\cite{ngtlib} and ANN-Benchmark~\cite{annbenchmark} for more details about the searching parameters.
On the other hand, SymQG adopts implicit re-ranking (see Section~\ref{subsec: greedy search}), which eliminates the re-ranking parameter.
Therefore, our method has smooth QPS-recall curves.

\begin{figure}[h]
    \centering
    \includegraphics[width=\linewidth]{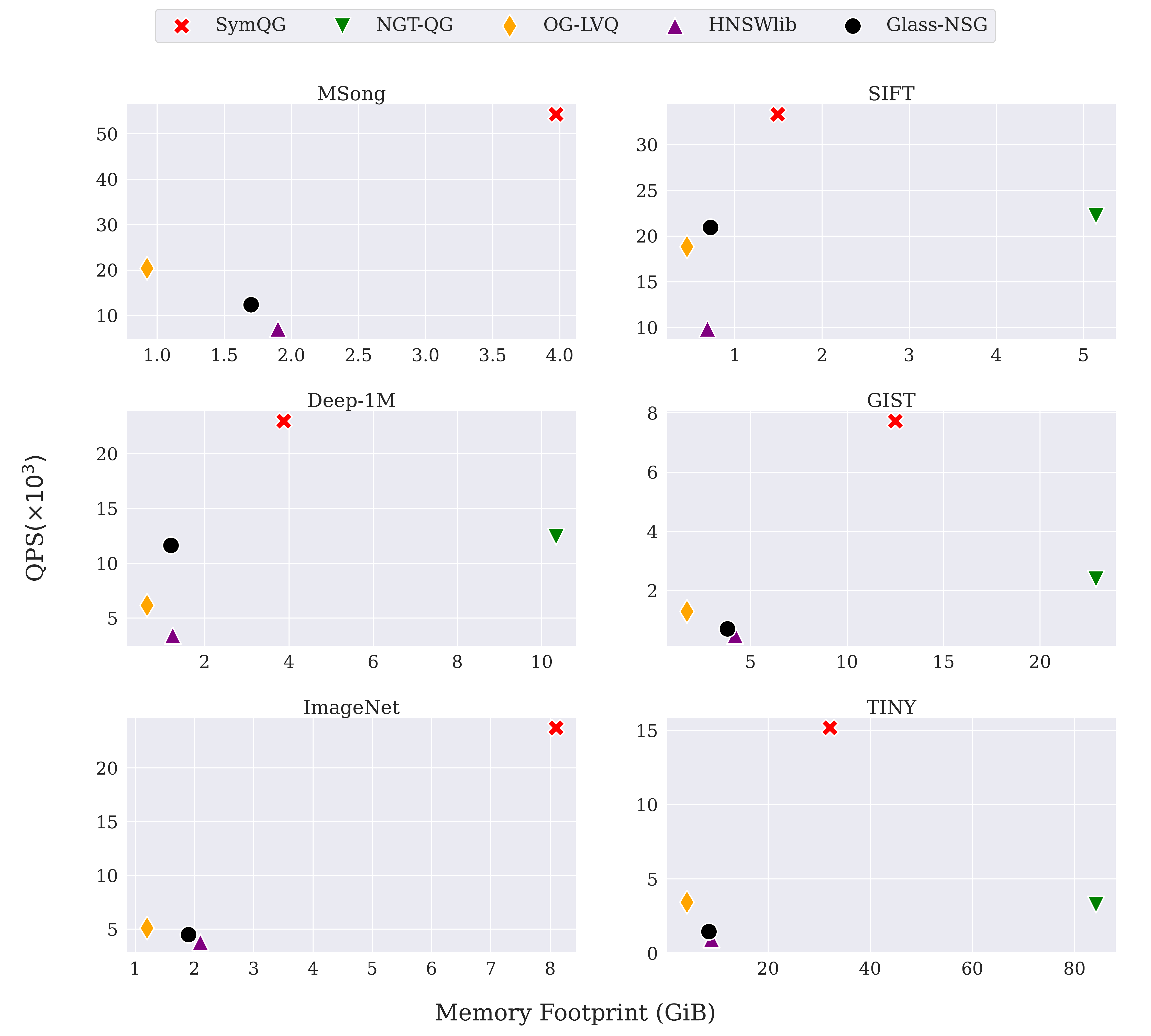}
    \vspace{-4mm}
    \caption{Trade-off between QPS and memory footprint for different methods (results with recall at 95\% and $K = 10$).
    The missing results of a certain method indicate that it fails to achieve at least 95\% recall on the dataset.}
    \vspace{-2mm}
    \label{fig:tradeoff}
\end{figure}

\smallskip\noindent\textbf{Indexing Time and Memory Footprint.}
We measure the indexing time and memory footprint for all the methods on all the datasets.
The indexing time is reported in Table~\ref{tab: indexing time}.
As illustrated, 
the indexing time of our method is comparable with HNSWlib and is much shorter than other methods.
The indexing time of NGT-QG is significantly higher than other methods.
This is because it adopts an extra process to optimize a constructed graph~\cite{onng}.
The memory footprint is reported in Table~\ref{tab: indexing size}.
As demonstrated, the memory footprint of our method and NGT-QG is higher than OG-LVQ, Glass-NSG and HNSWlib. 
This is because our method and NGT-QG search on graph-based indices with FastScan,
which requires to store multiple quantization codes for a 
single vertex (see Figure~\ref{fig:batch_graph} and Section~\ref{sec: intro}). 
This is likely to be unavoidable if we target to mitigate random memory accesses
during searching in pursuit of search performance.
In addition, compared with NGT-QG, the memory footprint of our method is much smaller,
partly because the quantization codes used by our method (which are produced by RaBitQ) are shorter than those used by NGT-QG (which are produced by PQ).
In Figure~\ref{fig:tradeoff}, we show time-space trade-off for different methods.
As illustrated, SymphonyQG achieves a significantly higher QPS compared with all baselines.
Although the space consumption of our method is higher than vanilla graph-based methods (i.e., HNSW and NSG),
it is still smaller than that of NGT-QG.
In summary, our method establishes the new state-of-the-art query performance and achieves a better time-space trade-off than NGT-QG.

\begin{figure}[h]
    \centering
    \includegraphics[width=\linewidth]{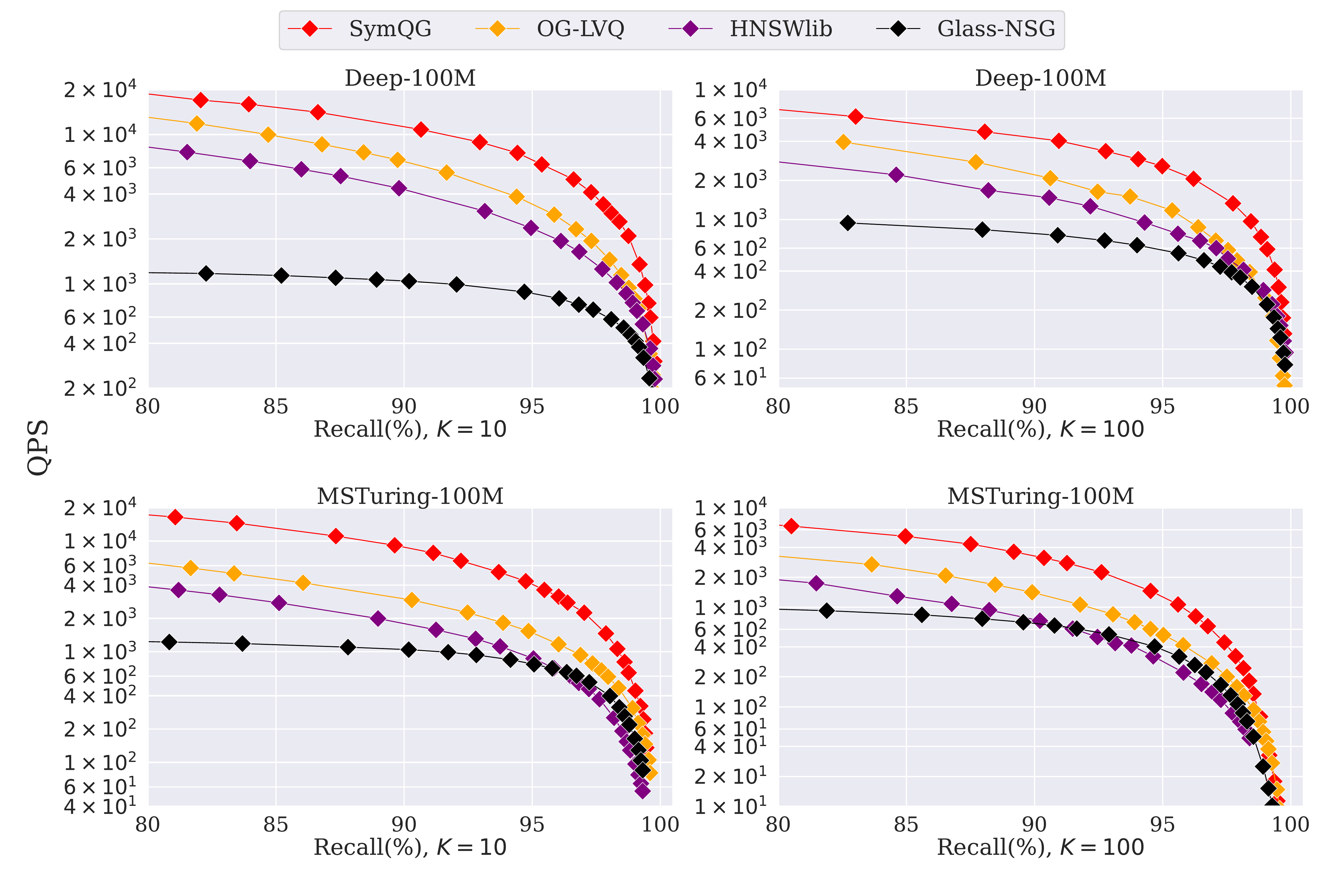}
    \vspace{-4mm}
    \caption{Evaluation of scalability.}
    \vspace{-2mm}
    \label{fig:large scale}
\end{figure}

\subsubsection{Scalability Study}
\label{subsec: scalability}
We verify the scalability of our method on Deep-100M and MSTuring-100M datasets.
The indexing parameters are set as $R=32$ and $EF=400$ for SymQG, OG-LVQ, Glass-NSG and HNSWlib.
We exclude NGT-QG in the evaluation in the scalability study
since NGT-QG fails to build an index with 1TB memory within 1 day (as a reference, our method can finish indexing within 2 hours).
We notice that there is another method named QBG, which is provided in the NGT library for handling large-scale datasets. 
However, as has been reported~\cite{intellvq}, the method cannot achieve reasonable recall (e.g., >85\%). 
Thus, we do not include QBG in the study either.
The QPS-recall curves are presented in Figure~\ref{fig:large scale},
and indexing time and size are shown in Table~\ref{tab: indexing time} and Table~\ref{tab: indexing size}, respectively.
The result shows that our method has good scalability and is still the most efficient on Deep-100M and MSTuring-100M.
Specifically, our method achieves at least 2.5x QPS compared with the most competitive baseline OG-LVQ at 95\% recall.
On the other hand, we note that the memory footprint of our method is larger than other methods by a clear margin. 
This is a limitation of our method caused by storing the data for FastScan (e.g., the quantization codes). 
We leave it as a future work to further optimize its memory consumption.

\begin{figure}[htbp]
    \centering
    \vspace{-2mm}
    \includegraphics[width=\linewidth]{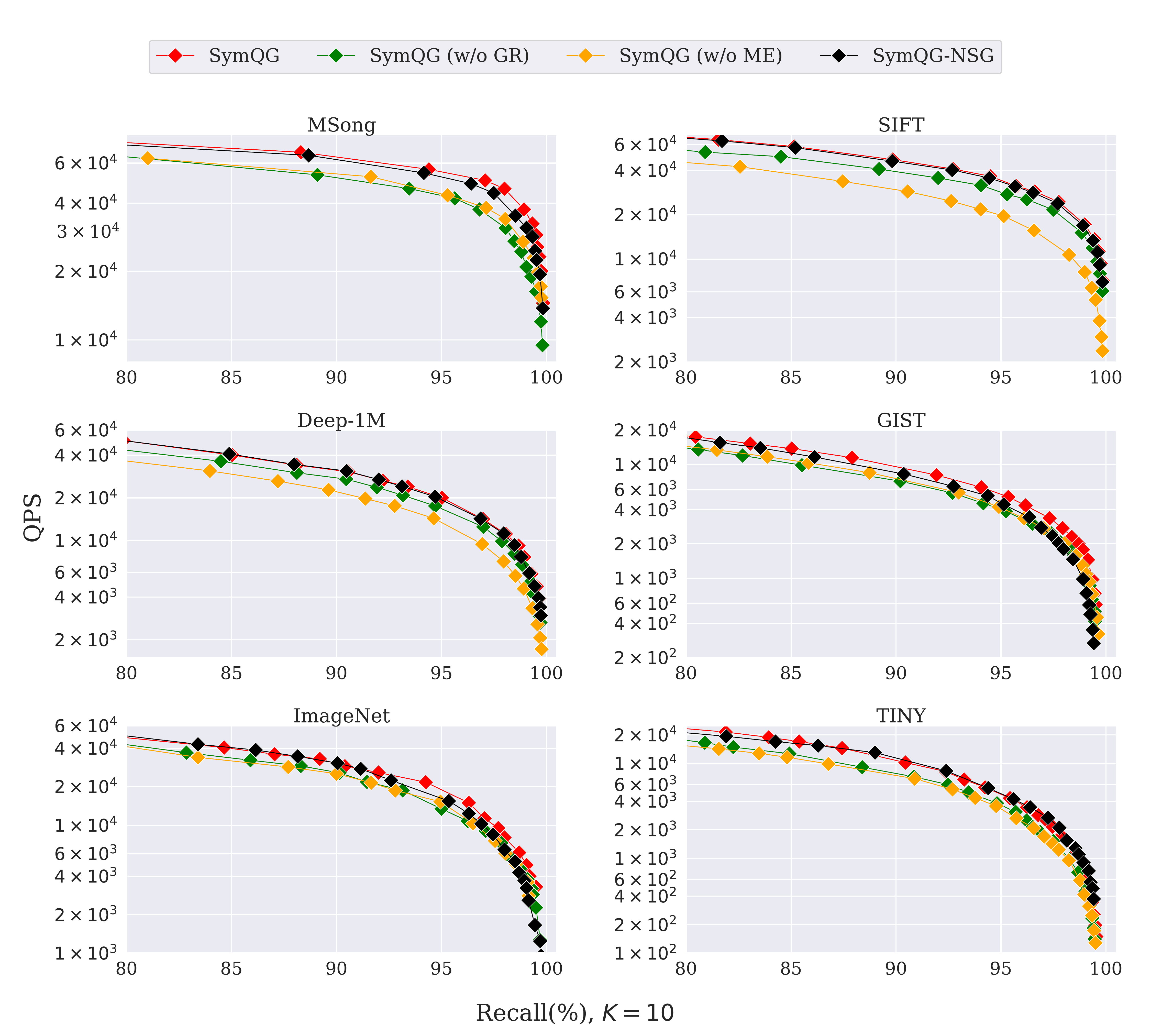}
    \vspace{-4mm}
    \caption{Ablation study on six moderate-scale datasets.
    }
    \vspace{-2mm}
    \label{fig: ablation}
\end{figure}

\subsubsection{Ablation Study}
\label{subsec: ablation}
In this section, we evaluate the effectiveness of the components in our method with an ablation study.
In particular, we study the effects of ablating a certain component from our method on the QPS-recall trade-off.
The study is conducted under the parameters of $R=32$, $EF=400$ and $K=10$
\footnote{ Due to the space limitation, we present more ablation study with different settings of $R$, $EF$ and $K$ in the 
appendix 
.}.
The ablation study involves the following three folds.

\smallskip\noindent\textbf{Evaluating the Impact of Using Multiple Estimated Distances on Query Performance.}
To show the effectiveness of using multiple estimated distances (Section~\ref{subsec: greedy search}),
we keep all other components while using only a single estimated distance for each vector during searching, 
i.e., we use the first estimated distance. We denote the method without using multiple estimated distances as SymQG (w/o ME).
According to Figure~\ref{fig: ablation}, 
SymQG surpasses SymQG (w/o ME) on all tested datasets,
which illustrates the effectiveness of using multiple estimated distances.

\smallskip\noindent\textbf{Evaluating the Impact of Our Efficient Indexing Algorithm on Index and Query Performance.}
We verify our index algorithm in both aspects of indexing and querying. 
(1) We use the original algorithm to build an NSG~\cite{nsg} ($R=32$, $EF=400$), and we apply all our other proposed techniques on it (including the graph refinement).
We prepare the data for FastScan on the constructed graph index. This method is denoted as SymQG-NSG.
We report the index time of SymQG-NSG and compare it with that of our method in Table~\ref{tab: indexing time of ablation}.
It shows that SymQG is at least 2.5x faster than SymQG-NSG in index construction.
(2) To test the quality (i.e., the search performance) of the graph constructed by our indexing algorithm (see Section~\ref{subsubec: efficient construction}),
 we test SymQG and SymQG-NSG on six moderate-scale datasets.
Figure~\ref{fig: ablation} shows that SymQG can always provide
competitive (or even slightly better) query performance on all tested datasets compared to SymQG-NSG. 
These results together indicate that our efficient indexing algorithm speeds up index construction significantly without harming the graph's quality in searching.

\begin{table}[!htbp]
    \centering
    \small
    \setlength{\tabcolsep}{0pt}
    \caption{Indexing time (minutes) for SymQG and SymQG-NSG. 
    For all datasets, $R=32$ and $EF=400$ during indexing.}
    \vspace{-4mm}
    \label{tab: indexing time of ablation}
    \begin{tabularx}{\columnwidth}{l *{6}{>{\centering\arraybackslash}X}}
    \toprule
        & MSong & SIFT & Deep-1M & GIST & ImageNet & TINY  \\
    \hline
        SymQG     & 0.8 & 0.3 & 0.6 & 1.7 & 1.1 & 4.7 \\
        SymQG-NSG & 3.0 & 1.7 & 2.3 & 6.2 & 3.7 & 22.5 \\
    \bottomrule
    \end{tabularx}
    \vspace{2mm}
\end{table}

\smallskip\noindent\textbf{Evaluating the Impact of the Graph Refinement Strategy on Query Performance.}
To test the effectiveness of the graph refinement strategy (Section~\ref{subsubsec: align fastscan}),
we build indices under same parameters while removing graph refinement process.
This is denoted as SymQG (w/o GR).
As presented in Figure~\ref{fig: ablation}, removing the component causes performance drop on all tested datasets.
We also report the average out-degree of the graph SymQG (w/o GR) in Table~\ref{tab: degrees of ablation}.
It shows that without the graph refinement strategy, 
the number of edges is indeed unaligned with the batch size of FastScan and thus, 
FastScan would waste much computation as the batch is not full. 
This helps to explain the performance drop of ablating the graph refinement component.

\subsubsection{Parameter Study}
\label{subsubsec: parameter study}

\begin{table}[!htbp]
    \centering
    \small
    \setlength{\tabcolsep}{0pt}
    \caption{Average out-degree of the graph without graph refinement. With graph refinement, each vertex's out-degree is guaranteed to be 32.}
    \vspace{-4mm}
    \label{tab: degrees of ablation}
    \begin{tabularx}{\columnwidth}{l *{6}{>{\centering\arraybackslash}X}}
    \toprule
        & MSong & SIFT & Deep-1M & GIST & ImageNet & TINY  \\
    \hline
        Degree & 19.8 & 25.9 & 29.2 & 18.1 & 17.9 & 19.7 \\
    \bottomrule
    \end{tabularx}
    \vspace{2mm}
\end{table}

\begin{figure}[htbp]
    \centering
    \vspace{-2mm}
    \includegraphics[width=\linewidth]{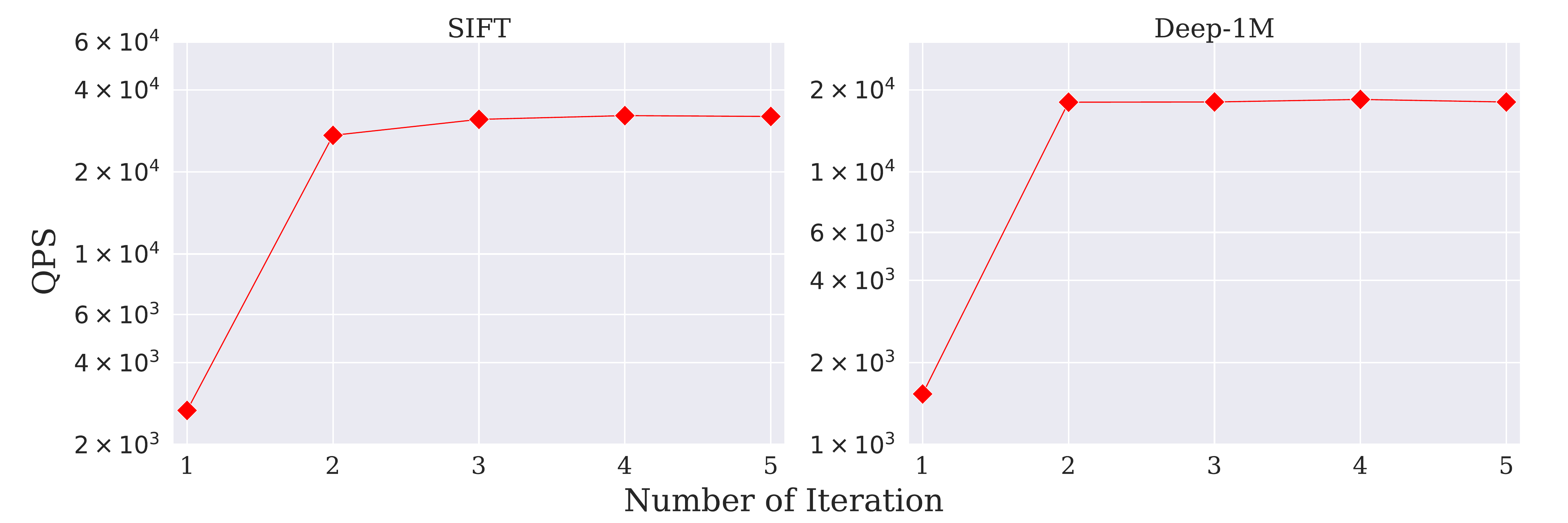}
    \vspace{-4mm}
    \caption{QPS at 95\% recall ($K=10$) for different number of iterations during indexing of SymQG.
    }
    \vspace{-2mm}
    \label{fig: iter ablation}
\end{figure}

In our algorithm, we introduce a new parameter in indexing, i.e., the number of iterations of Algorithm~\ref{alg:indexing}.
To illustrate its influence on the quality of the graph for ANN queries,
we build indices with different iterations and test their query performance.
As presented in Figure~\ref{fig: iter ablation},
when the number of iterations is 3 and larger, the search performance keeps stable on two different datasets.
Therefore, we  would like to suggest to set the number of iterations in graph construction to be 3 or 4
for our method by default.

\section{Related Work}
\label{sec: related}

Existing algorithms for ANN search in high-dimensional vectors can be roughly classified into four categories: 
the graph-based~\cite{fanng, nsw, fu2016efanna, iwasaki2016pruned, spann, rng, hnsw,nsg,2019diskann, SPTAG, nssg, tau-mng, azizi2023elpis}, 
the quantization-based~\cite{hakanvq,pq, pqfs, opq, addq, guo2020accelerating, babenko2014inverted},
the hashing-based~\cite{gan2012locality, datar2004locality, huang2015query, indyk1998approximate, sun2014srs, tao2010efficient}
and the tree-based~\cite{beygelzimer2006cover, arya1993approximate, gu2022parallel, mtree}.
Graph-based methods offer the state-of-the-art time-accuracy trade-off on real-world datasets~\cite{annbenchmark, bigann2021, bigann2023}.
Quantization-based methods compress high-dimensional vectors into short quantization codes~\cite{pq, opq, addq, guo2020accelerating, babenko2014inverted} and use the compressed codes to estimate distances.
FastScan~\cite{pqfs, quickadc} significantly accelerates the distance estimation of these methods, 
making them an important component in many in-memory ANN libraries~\cite{guo2020accelerating, faiss,ngtlib}.
Hashing-based methods promise a theoretical guarantee on the probability of finding $c$-approximate nearest neighbors.
However, they often struggle to achieve high recall with competitive efficiency in practice.
Tree-based methods are powerful on low-dimensional datasets but their search performance suffers 
from the curse of dimensionality in the high-dimensional scenarios.
For more comprehensive understanding of different ANN methods,
we refer readers to recent tutorials and surveys~\cite{tutorialThemis, tutorialXiao, li2019approximate, dobson2023scaling_billion_benchmark, wang2023graph, aumullerrecent2023, wang_survey, graph_benchmark,marco_survey,marco_graph}.

Moreover, since graph-based methods offer the state-of-the-art query performance,
some existing studies target to optimize certain aspects of indexing and querying for graph-based methods.
For example, $\tau$-MNG~\cite{tau-mng}, NSSG~\cite{nssg} and ONNG~\cite{onng} try to fine-tune the graph structure 
to obtain a better query performance.
FINGER~\cite{chen2023finger} and ADS{\small AMPLING}~\cite{adsampling} target to reduce the cost of distance computation.
Vamana~\cite{2019diskann} and OG-LVQ~\cite{intellvq} focus on decreasing the memory footprint of graph-based indices on large-scale datasets. 
ELPIS~\cite{azizi2023elpis} aims at reducing the construction time of graph-based indices.
In this paper, we aim to integrate quantization and graph-based indices more symphoniously and leave it as future work to apply and/or adapt these optimization techniques to boost our method.

Besides the ANN query in the Euclidean space, we note that there is a vast literature which concerns the similarity 
search question for other data types and {similarity metrics~\cite{echihabi2023pros, echihabi2022hercules, marco_SISAP, marco_survey, moromips, zhoumips}}.
We refer readers to surveys and reference books on the similarity search question in 
more generic metric spaces~\cite{marco_survey, similarity_search_book1, similarity_search_book2}.
Our library focus on the ANN query in the Euclidean space (including the metrics of Euclidean distances and cosine similarity) 
while its design might provide some insights for other similarity search questions. 
For example, graph-based indices have also been used 
for maximum inner product search~\cite{moromips,zhoumips}.
It is also possible to enhance these methods with similar ideas of this paper, e.g., conducting re-ranking implicitly and using multiple estimated distances during graph-based searching.
However, we note that developing a library for other similarity search questions involves highly non-trivial challenges and workloads, which is out of the scope of this paper. 
We notice that recently, a thread of studies show the potential of applying the techniques in the similarity search question of data series to that of {high-dimensional vectors~\cite{echihabi2023pros,echihabi2022hercules}}.
It remains to be an interesting question whether a well-optimized library based on these techniques can outperform the libraries based on graph-based indices.

\section{Conclusion and Discussion}
\label{sec: conclusion}

In conclusion, in this paper,
we propose a new method named SymphonyQG which targets to fully unleash the 
potential of integrating quantization and graph-based indices in a symphonious way.
For querying, we incorporate RaBitQ~\cite{gao2024rabitq} into graph-based indices for its superiority over the conventional PQ method.
Moreover, we design a novel search algorithm, 
by which our method can eliminate the random memory accesses caused by an explicit re-ranking step. 
In this search algorithm, we propose to use multiple estimated distances in graph-based searching to boost the accuracy.
For indexing, we use our efficient search algorithm to accelerate graph index construction.
We further propose to refine the graph structure of SymphonyQG such that it is better
aligned with the in-batch computations of FastScan.
Based on all the proposed techniques, SymphonyQG establishes the new state-of-the-art in terms of time-accuracy 
trade-off as well as significantly speeding up the indexing construction.

On the other hand, our method still has certain limitations. 
Specifically, as discussed in Section~\ref{sec: intro} and Section~\ref{sec: experiments}, 
SymphonyQG stores duplicated quantization codes for a single vector,
which increases the space consumption.
Therefore, our method may be unsuitable for memory-limited scenarios (e.g., cloud functions~\cite{vexless2024}).
Also, SymphonyQG can handle ANN query with the metrics of Euclidean distance and cosine similarity only
since the quantization method we use (i.e., RaBitQ) is designed for these metrics.
More efforts are needed to extend (some of) our techniques for other methods.
For future work, we would like to discuss the following possible directions.
(1) SymphonyQG has promising performance in indexing. This implies its potential of supporting efficient dynamic updates. 
For example, for deleting a vertex from the graph, it suffices to re-build the edges of the vertices which have an edge towards the vertex. 
This can be achieved by running the indexing algorithm for these vertices.
(2) The relatively large index size of SymphonyQG limits its application scenarios.
Therefore, compression algorithms (e.g., dimension reduction) can be used to reduce its memory consumption.

\section*{Acknowledgements}
This research is supported by the Ministry of Education, Singapore, under its Academic Research Fund (Tier 2 Award MOE-T2EP20221-0013, Tier 2 Award MOE-T2EP20220-0011, and Tier 1 Award (RG20/24)). Any opinions, findings and conclusions or recommendations expressed in this material are those of the author(s) and do not reflect the views of the Ministry of Education, Singapore.
We would like to thank the anonymous reviewers for providing constructive feedback and valuable suggestions.
Additionally, we would like to thank Masajiro Iwasaki, a researcher from Yahoo Japan, for patiently explaining the details of NGT-QG to us.

\bibliographystyle{ACM-Reference-Format}
\bibliography{main}

\appendix
\section*{appendix}

\section{Indexing Parameters for Different Methods}
\label{sec: index param}
In Section~\ref{subsec: exp moderate scale}, we choose optimal (highest QPS at 95\% call, $K=10$) indexing parameters for different methods to test their query performance.
The details are presented in Table~\ref{tab: indexing param}.
Here, we omit common parameters ($EF=400$) for SymQG, OG-LVQ, Glass-NSG and HNSWlib.
The indexing of NGT-QG involves many parameters, and we only show the group name of the chosen index. 
For detailed information, we refer readers to the settings provided in ANN-Benchmark~\cite{annbenchmark}.

\begin{table}[!htbp]
    \centering
    \caption{Indexing parameters for different methods.}
    \label{tab: indexing param}
    \vspace{-4mm}
    \small
    \setlength{\tabcolsep}{0pt}
    \begin{tabularx}{\columnwidth}{l *{6}{>{\centering\arraybackslash}X}}
    \toprule
        & SymQG & NGT-QG & OG-LVQ & Glass-NSG & HNSWlib \\
    \hline
        MSong       & $R=32$ & s20000-e0.1 & $R=32$, LVQ-8x8 & $R=32$, level0 & $M=32$ \\
        SIFT        & $R=32$ & s4000-e0.08 & $R=64$, LVQ-4x8 & $R=64$, level2 & $M=16$ \\
        Deep-1M     & $R=64$ & s20000-e0.1 & $R=64$, LVQ-4x8 & $R=64$, level2 & $M=32$ \\
        GIST        & $R=64$ & s20000-e0.1 & $R=64$, LVQ-4x8 & $R=64$, level0 & $M=64$ \\
        ImageNet    & $R=64$ & s20000-e0.1 & $R=64$, LVQ-8x8 & $R=64$, level0 & $M=32$ \\
        TINY        & $R=64$ & s20000-e0.1 & $R=64$, LVQ-4x8 & $R=64$, level0 & $M=32$ \\
        Deep-100M   & $R=32$ & - & $R=32$, LVQ-8x8 & $R=32$, level0 & $M=16$ \\
        MSTuring-100M   & $R=32$ & - & $R=32$, LVQ-8x8 & $R=32$, level2 & $M=16$ \\
    \bottomrule
    \end{tabularx}
    \vspace{-2mm}
\end{table}

\section{Ablation Study of Proposed Techniques}
\label{sec: more ablation}
\begin{figure}[htbp]
    \centering
    \begin{subfigure}[b]{0.9\linewidth}
         \centering
         \vspace{-2mm}
         \includegraphics[width=\linewidth]{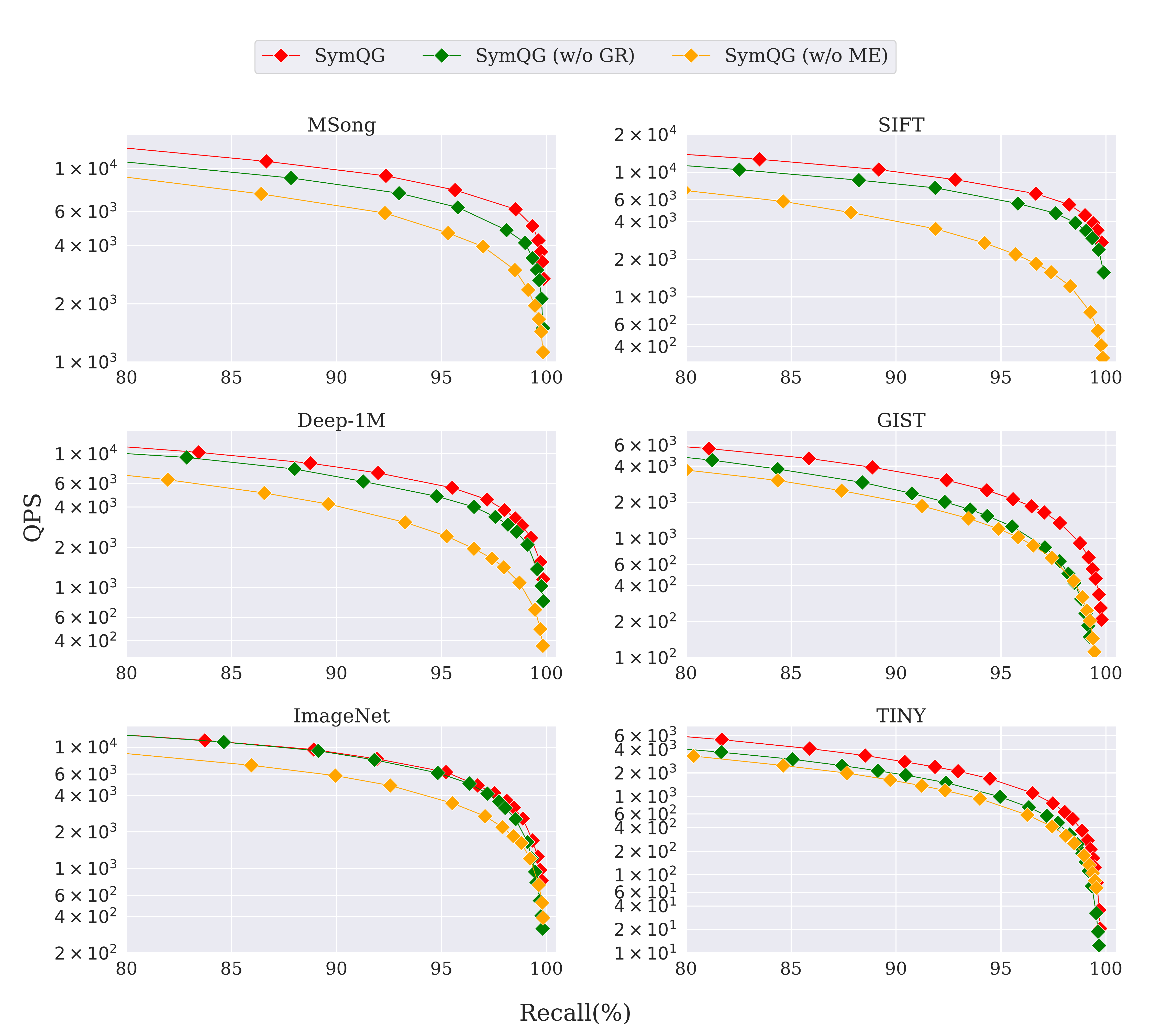}
         \vspace{-4mm}
         \caption{$K=100,R=32,EF=400$}
    \end{subfigure}
    \hfill
    \begin{subfigure}[b]{0.9\linewidth}
        \centering
        \includegraphics[width=\linewidth]{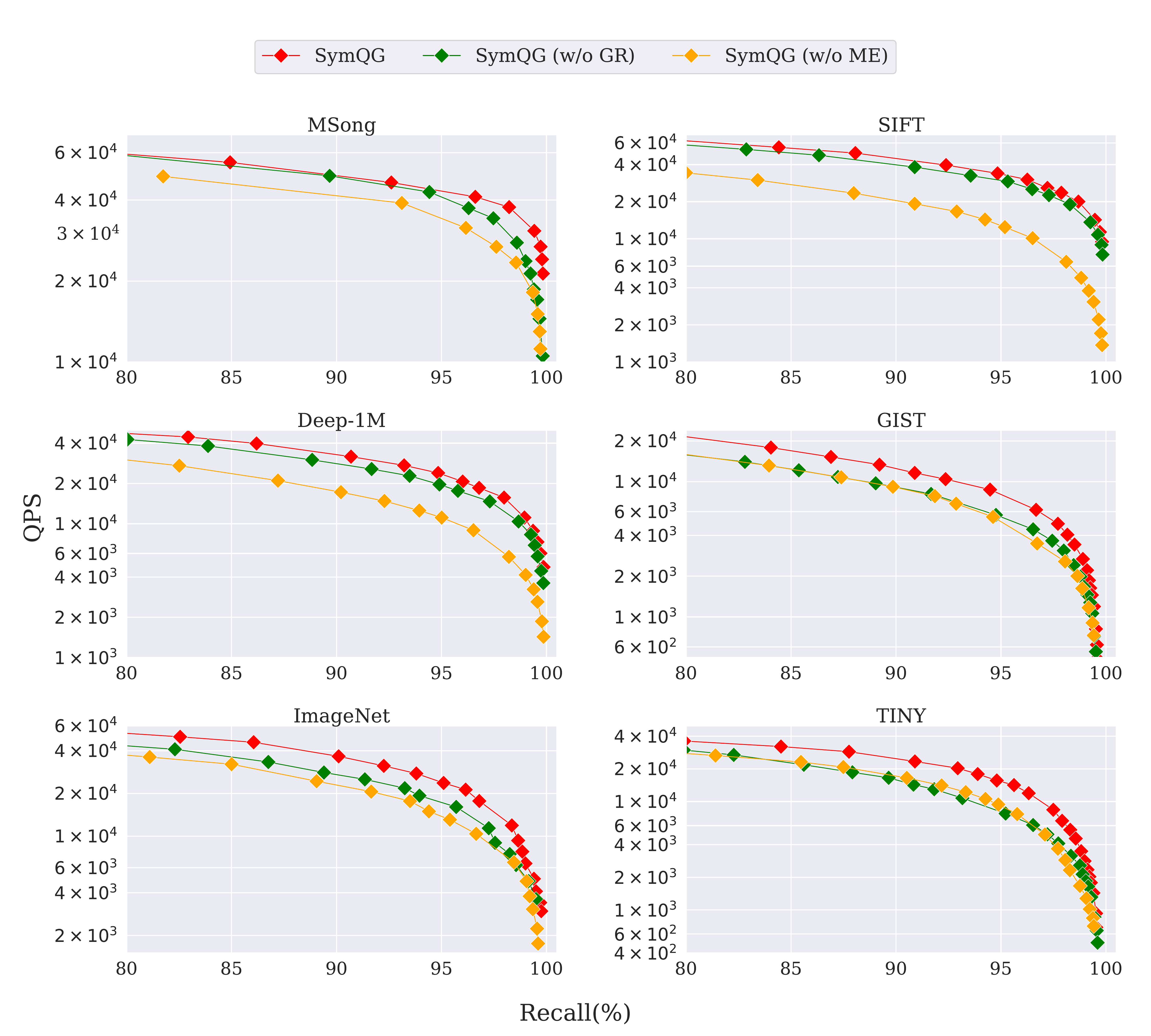}
        \vspace{-4mm}
        \caption{$K=10,R=64,EF=400$}
     \end{subfigure}
     \hfill
     \begin{subfigure}[b]{0.9\linewidth}
        \centering
        \includegraphics[width=\linewidth]{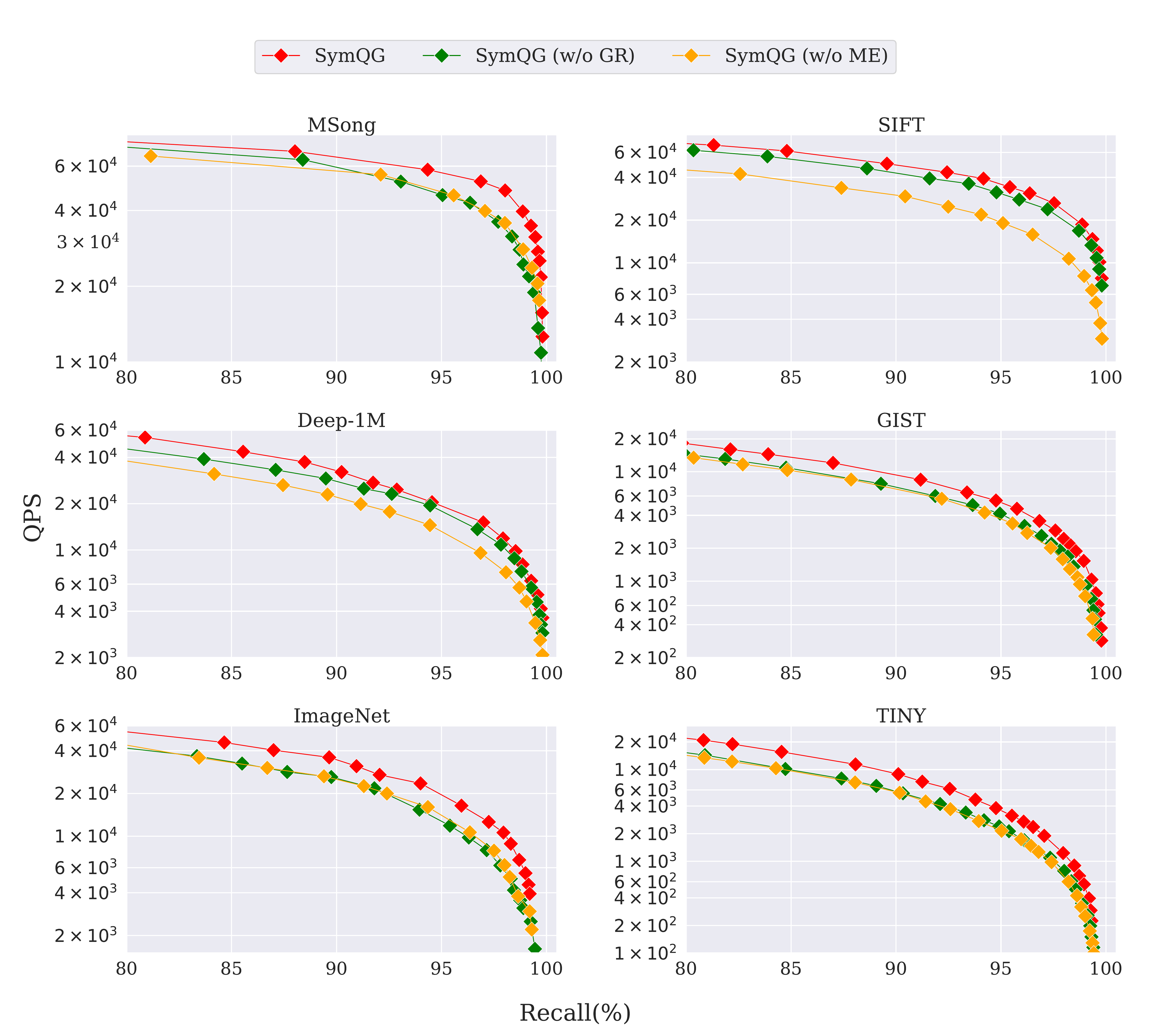}
        \vspace{-4mm}
        \caption{$K=10,R=32,EF=200$}
    \end{subfigure}
    \vspace{-2mm}
        \caption{Ablation study on six moderate-scale datasets with different settings.}
    \label{fig:more ablation}
\end{figure}

In Section~\ref{subsec: ablation}, 
we study the effectiveness of two proposed techniques (i.e., using multiple estimated distances during querying and the graph refinement strategy)
by setting $R=32$, $EF=400$ for indexing and $K=10$ for searching.
To further evaluate the effectiveness and generality of our proposed techniques,
we test the performance gain of 
the proposed techniques
with respect to the following parameters:
(1) $EF$, the parameter which controls the edge construction during graph-based indexing, 
(2) $R$, the maximum out-degree and (3) $K$, the number of nearest neighbors we retrieve.
To be more specific,
we changed $EF$, $R$ and $K$ based on the setting we used in Section~\ref{subsec: ablation}
and tested the query performance of SymQG, SymQG(w/o ME) and SymQG(w/o GR) on six moderate-scale datasets.
The detailed settings and experimental results are presented in Figure~\ref{fig:more ablation}.
As illustrated, 
SymQG surpasses its variants with different settings on all tested datasets,
which further  verifies the effectiveness of using multiple estimated distances and graph refinement strategy.

\end{document}